\documentclass[aps,prl,twocolumn,superscriptaddress,amsmath,amssymb]{revtex4-1}
\usepackage{graphicx}
\usepackage{dcolumn,tabularx}
\usepackage{bm}
\usepackage{braket}

\usepackage{mathdots}
\usepackage[T1]{fontenc}
\usepackage[unicode=true]{hyperref}
\usepackage{booktabs}
\usepackage{mathtools}
\usepackage{color}
\usepackage{dsfont}
\usepackage{mathrsfs}
\usepackage{babel,calc,amsmath,amsthm,amssymb,graphicx,subfigure,xcolor,comment}
\usepackage{mathdots}
\usepackage[T1]{fontenc}
\usepackage[unicode=true]{hyperref}
\usepackage{braket}
\usepackage{dsfont}
\usepackage[normalem]{ulem}
\hypersetup{
	colorlinks=true,       		
	linkcolor=blue,          	
	citecolor=red,             
	urlcolor=magenta,          
}

\newcommand{\fxy}[1]{{\color{blue}#1}}

\newcommand{\abs}[1]{\left\vert#1\right\vert}

\newcommand{\exv}[1]{\left\langle#1\right\rangle}

\newif\ifdebug

\debugtrue

\ifdebug
\definecolor{zhliu}{rgb}{0.5, 0.03, 0}

\newcommand{\note}[1]{\textcolor{zhliu}{#1}}
\newcommand\delete{\bgroup\markoverwith{\textcolor{zhliu}{\rule[0.5ex]{2pt}{0.8pt}}}\ULon}

\else

\newcommand{\note}[1]{\ignorespaces}
\newcommand{\delete}[1]{\ignorespaces}

\fi

\begin{document}
	\renewcommand{\figureautorefname}{Fig.}
	\renewcommand{\figurename}{Fig.}
	\title{Observing tight triple uncertainty relations in two-qubit systems}
	
	\affiliation{CAS Key Laboratory of Quantum Information, University of Science and Technology of China, Hefei 230026, China}
	\affiliation{Anhui Province Key Laboratory of Quantum Network, University of Science and Technology of China, Hefei 230026, China}
	\affiliation{CAS Center for Excellence in Quantum Information and Quantum Physics, University of Science and Technology of China, Hefei 230026, China}

	\author{Yan Wang}
	\affiliation{CAS Key Laboratory of Quantum Information, University of Science and Technology of China, Hefei 230026, China}
	\affiliation{Anhui Province Key Laboratory of Quantum Network, University of Science and Technology of China, Hefei 230026, China}
	\affiliation{CAS Center for Excellence in Quantum Information and Quantum Physics, University of Science and Technology of China, Hefei 230026, China}
	\affiliation{School of Physics, Hangzhou Normal University, Hangzhou 310036, China}

	\author{Jie~Zhou}
	\affiliation{College of Physics and Materials Science, Tianjin Normal University, Tianjin 300382, China}
	
	\author{Xing-Yan Fan}
	\affiliation{Theoretical Physics Division, Chern Institute of Mathematics, Nankai University, Tianjin 300071, China}
	
	\author{Ze-Yan Hao}
	\affiliation{CAS Key Laboratory of Quantum Information, University of Science and Technology of China, Hefei 230026, China}
	\affiliation{Anhui Province Key Laboratory of Quantum Network, University of Science and Technology of China, Hefei 230026, China}
	\affiliation{CAS Center for Excellence in Quantum Information and Quantum Physics, University of Science and Technology of China, Hefei 230026, China}
	
	\author{Jia-Kun Li}
	\affiliation{CAS Key Laboratory of Quantum Information, University of Science and Technology of China, Hefei 230026, China}
	\affiliation{Anhui Province Key Laboratory of Quantum Network, University of Science and Technology of China, Hefei 230026, China}
	\affiliation{CAS Center for Excellence in Quantum Information and Quantum Physics, University of Science and Technology of China, Hefei 230026, China}
	
	\author{Zheng-Hao Liu}
	\affiliation{CAS Key Laboratory of Quantum Information, University of Science and Technology of China, Hefei 230026, China}
	\affiliation{Anhui Province Key Laboratory of Quantum Network, University of Science and Technology of China, Hefei 230026, China}
	\affiliation{CAS Center for Excellence in Quantum Information and Quantum Physics, University of Science and Technology of China, Hefei 230026, China}
	
	\author{Kai Sun}
	\email{ksun678@ustc.edu.cn}
	\affiliation{CAS Key Laboratory of Quantum Information, University of Science and Technology of China, Hefei 230026, China}
	\affiliation{Anhui Province Key Laboratory of Quantum Network, University of Science and Technology of China, Hefei 230026, China}
	\affiliation{CAS Center for Excellence in Quantum Information and Quantum Physics, University of Science and Technology of China, Hefei 230026, China}

	\author{Jin-Shi Xu}
	\affiliation{CAS Key Laboratory of Quantum Information, University of Science and Technology of China, Hefei 230026, China}
	\affiliation{Anhui Province Key Laboratory of Quantum Network, University of Science and Technology of China, Hefei 230026, China}
	\affiliation{CAS Center for Excellence in Quantum Information and Quantum Physics, University of Science and Technology of China, Hefei 230026, China}
	\affiliation{Hefei National Laboratory, University of Science and Technology of China, Hefei 230088, China}
	
	\author{Jing-Ling~Chen}
	\email{chenjl@nankai.edu.cn}
	\affiliation{Theoretical Physics Division, Chern Institute of Mathematics, Nankai University, Tianjin 300071, China}

	\author{Chuan-Feng~Li}
	\email{cfli@ustc.edu.cn}
	\affiliation{CAS Key Laboratory of Quantum Information, University of Science and Technology of China, Hefei 230026, China}
	\affiliation{Anhui Province Key Laboratory of Quantum Network, University of Science and Technology of China, Hefei 230026, China}
	\affiliation{CAS Center for Excellence in Quantum Information and Quantum Physics, University of Science and Technology of China, Hefei 230026, China}
	\affiliation{Hefei National Laboratory, University of Science and Technology of China, Hefei 230088, China}

	\author{Guang-Can Guo}
	\affiliation{CAS Key Laboratory of Quantum Information, University of Science and Technology of China, Hefei 230026, China}
	\affiliation{Anhui Province Key Laboratory of Quantum Network, University of Science and Technology of China, Hefei 230026, China}
	\affiliation{CAS Center for Excellence in Quantum Information and Quantum Physics, University of Science and Technology of China, Hefei 230026, China}
	\affiliation{Hefei National Laboratory, University of Science and Technology of China, Hefei 230088, China}

	\begin{abstract}		
		As the fundamental tool in quantum information science, the uncertainty principle is essential for manifesting nonclassical properties of quantum systems.
		Plenty of efforts on the uncertainty principle with two observables have been achieved, making it an appealing challenge to extend the scenario to multiple observables. 
		Here, based on an optical setup, we demonstrate the uncertainty relations in two-qubit systems involving three physical components with the tight constant $2/\sqrt{3}$, which signifies a more precise limit in the measurement of multiple quantum components and offers deeper insights into the trade-offs between observables. 
		Furthermore, we reveal the correspondence of the maximal values of the uncertainty functions and the degree of entanglement, where the more uncertainty is proportional to the higher degree of entanglement. 
		Our results provide a new insight into understanding the uncertainty relations with multiple observables and may motivate more innovative applications in quantum information science.
	\end{abstract}
	
	\maketitle
	The uncertainty principle, describing the inherent incompatibility of quantum measurement, is one of the most fundamental laws in quantum mechanics \cite{heisenberg1927anschaulichen,1927Kennard,1928Weyl}.
	Through different variance-based measures of two physical quantities, e.g., energy and time, angular momentum and angle, and so on, various uncertainty relations in terms of the product or additive forms have been demonstrated \cite{1957Hirschman,1975Beckner,1975Birula,1983PRLDeutsch,1987PRDKraus,1988PRLMaassen,1996APBraunstein,1998PLARuiz,2003Ghirardi,2006PRABirula,2010NJPWehner,2011PRAPartovi,2017PMPColes}.
	Typically, the Heisenberg-Robertson uncertainty relation \cite{1929PRRobertson}, also regarded as the preparation uncertainty, is associated with incompatible measurement outcomes of two arbitrary observables for some state preparations.  
	
	In comparison to two-observable uncertainty relations like Heisenberg's, the observables are further extended to multiple ($\ge$ three) cases from different perspectives \cite{2014PRAKechrimparis,liang2011specker,PhysRevA.99.032107,PhysRevA.98.032118,2017PRLfei,wang2023uncertainty,PhysRevLett.131.150203}. 
	For example, a tight and non-trivial form of uncertainty relation for the triple components of angular momentum $\Vec{S}=(S_x, S_y, S_z)$ was derived as $\Delta S_x \Delta S_y \Delta S_z\geq  |(\lambda^3/8)\langle S_x \rangle\langle S_y \rangle\langle S_z \rangle|^{1/2}$ and experimentally demonstrated by Ma and his collaborators \cite{2017PRLfei}, which is involving with an additional parameter $\lambda=2/\sqrt{3}$ and $\Delta O\equiv \sqrt{\langle O^2 \rangle-\langle O\rangle^2}$ depicting the standard variation of the observable $O$ corresponding to a known state $\rho$. Remarkably, it gives clear physical meanings to each of the quantities above (i.e., the expectation or standard deviation relating to the components of angular momentum operator $\vec{S}$), and takes the equal sign with a nonzero value on both sides \cite{2017PRLfei}. 
	However, due to the simplified structure and physical limitations in the single qubit system, it is of high significance to investigate the multi-observable uncertainty relations based on the two- and multi-qubit systems in harnessing the full computational power, error correction capabilities, and entanglement properties required for advanced the study of multiple incompatible quantum measurements in quantum computing and information processing \cite{ladd2010quantum,georgescu2014quantum,RevModPhys.95.011003}.
	
	Moreover, the numerous applications based on uncertainty relations have been investigated in quantum information science \cite{RevModPhys.95.011003}, including quantum metrology \cite{giovannetti2011advances}, quantum random number generation \cite{PhysRevA.90.052327,PhysRevX.6.011020} and entanglement detection \cite{RevModPhys.81.865}. 
	Particularly, the entropic uncertainty relation has been demonstrated in witnessing entanglement theoretically and experimentally \cite{li2011experimental,prevedel2011experimental,PhysRevA.107.052617}. 
	Here, definite forms of uncertainty relations with physical quantities components in the two-qubit systems have been built to explore extensive relations between uncertainty and entanglement, underscoring fundamental insights into the essence of quantum measurement limits.
	
	In this paper, we experimentally demonstrate the tight triple uncertainty relations in the two-qubit systems with three explicit physical observables. 
	Specifically, the tight constant $2/\sqrt{3}$ has been demonstrated universally applicable across different perspectives of uncertain relations. 
	The attainable and nontrivial bound of the triple uncertainty relation is experimentally observed through preparing the separable quantum state.
	Besides, we prepare a series of quantum states to verify the associations between the maximal values of the uncertainty functions and the concurrence of the two-qubit states, including pure and mixed states, which shows the tight triple uncertainty relations could also be a useful tool for the entanglement identification in two-qubit systems. Our results enrich the understanding of uncertainty principles in two-qubit systems and make contributions to exploring more in-depth advancements in physical quantum information processing, entanglement detection, and the development of optimal quantum measurement techniques \cite{huang2023entropic}.
	
	\begin{figure}[t]
		\includegraphics[width=0.5\textwidth]{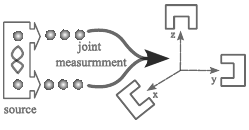}
		\caption{\textbf{The theoretical framework.} The particle pairs from the two-qubit source are sent to joint measurements, consisting of three mutually perpendicular measurement directions denoting $x$, $y$\fxy, and $z$. The triple uncertainty relationship means the incompatible measurement outcomes among the three measuring observables.
		}\label{th}
	\end{figure}
	
	\emph{The uncertainty relationships of two-qubit system.} As shown in Fig. \ref{th}, the two-qubit source emits copies of particle pairs.
	For each copy pair, three mutually perpendicular measurement directions ($x$, $y$ and $z$) are chosen and realized through different combinations of joint measurements. 
	Notably, we never attempt to measure all directions simultaneously for the same particle.
	The triple uncertainty relation states it is impossible to predict the outcomes of the measurement directions $x$, $y$ and $z$ simultaneously \cite{2014PRAKechrimparis,2017PRLfei}.
	More precisely, when one of the measurement outcomes of the three directions becomes better, the outcomes of the left two would be more uncertain \cite{2017PMPColes}.
	The core point of building the triple uncertainty relations is to give the physical forms of the above three measurement directions.
	Here, we explain the triple uncertainty relations with the directions $x$, $y$, and $z$ representing the three components of the operator $\vec{J}$ and the total angular momentum $\vec{K}$ which are built on the two-qubit systems, and discuss the tight forms in the product and additive ways.

	The operator $\vec{J}$ is defined by cross product form as $\vec{J}=\vec{S}_1\times \vec{S}_2$ with the angular momentum for $i$-th qubit as $\vec{S}_i=(S_{ix}, S_{iy}, S_{iz})$ $(i=1,2)$, which serves as a shift operator in Heisenberg chain models~\cite{PhysRevE.60.1486} and is related to the important Dzyaloshinsky-Moriya (DM) interaction in condensed matter theory~\cite{dzyaloshinsky1958thermodynamic,PhysRevLett.4.228}. 
	And the three components of $\vec{J}$ could be written as $J_i =\sum_{j,k} \epsilon_{ijk} S_{1j} S_{2k}\ (i,j,k=x,y,z)$, where $\epsilon_{ijk}$ indicates the Levi-Civita symbol.
	By multiplying the standard deviations of the above three components, the corresponding uncertainty inequality is $\Delta J_x\Delta J_y\Delta J_z\ge|(1/8)\langle R_x \rangle \langle R_y \rangle \langle R_z\rangle|^{1/2}$, with $R_l=(1/4)(S_{1l}+S_{2l})\ (l=x,y,z)$.
	To find the lower bound of the above inequality, we introduce the triple constant $\lambda$ and rewrite the tight triple uncertainty relation as $\Delta J_x\Delta J_y\Delta J_z\ge|(\lambda^3/8)\langle R_x \rangle \langle R_y \rangle \langle R_z\rangle|^{1/2}$.
	Through mathematical minimizing, the triple constant is obtained as $\lambda=2/\sqrt{3}$, and the separable quantum state $\rho_{12}$ saturates the bound of the product form of the triple uncertainty relation.
	More calculation details are listed in S1 of the Supplementary Materials (SM)~\cite{urSM}.	
	
	The total angular momentum operator defined as $\vec{K}=\vec{S}_1\otimes\mathds{1}+\mathds{1}\otimes \vec{S}_{2}=4\,\vec{R}$, with the components as $K_l=S_{1l}\otimes\mathds{1}+\mathds{1}\otimes S_{2l}$ $(l=x,y,z)$ with the two-dimensional identity operator $\mathds{1}$ \cite{sakurai2020modern}. Similarly, the tight triple uncertainty relation concerning the angular momentum $\vec{K}$ is written in $\Delta K_x\Delta K_y\Delta K_z\ge|(\lambda^3/8)\langle {K}_x \rangle \langle {K}_y \rangle \langle {K}_z\rangle|^{1/2}$, which could also be optimized, with $\lambda=2/\sqrt{3}$ (See more details in S2 of SM~\cite{urSM}).
	
	Based on the above tight triple uncertainty relations in the product and additive forms for the operator $\vec{J}$ and the total angular momentum $\vec{K}$, we can further construct the uncertainty functions as
	\begin{align}\label{eq:ent-0}
		f&=\Delta \widetilde{J_x} \Delta \widetilde{J_y} \Delta \widetilde{J_z}
		- \left\lvert\frac{\lambda^3}{8}\langle\widetilde{R_x}\rangle \langle \widetilde{R_y}\rangle \langle \widetilde{R_z}\rangle\right\lvert^{1/2}\geq 0,\\
		g&=\sum_k (\Delta\widetilde{J_k})^2-\frac{\lambda}{2}(\sum_k|\langle\widetilde{R_k}\rangle|)\geq 0, \\
		h&=\Delta \widetilde{K_x} \Delta \widetilde{K_y} \Delta \widetilde{K_z}
		- \left\lvert\frac{\lambda^3}{8}\langle\widetilde{K_x} \rangle\langle \widetilde{K_y}\rangle \langle \widetilde{K_z}\rangle\right\lvert^{1/2}\geq 0,\\
		k&=\sum_l \left(\Delta\widetilde{K_l}\right)^2-\frac{\lambda}{2}\left(\sum_l|\langle\widetilde{K_l}\rangle|\right)\geq 0,
		\label{eq:ent-1}
	\end{align}
	where $(k,l=x,y,z)$, $\lambda$ is equal to $ 2/\sqrt{3}$,
	and the operator $\widetilde{O_i}=UO_iU^\dagger$ $(i=x,y,z)$ with the local unitary operator $U$ applied to obtain the four maximal values of the uncertainty functions $f$, $h$, $g$ and $k$.

	The above four uncertainty functions could be associated with the concurrence of two-qubit systems. For any Bell-type states like $\ket{\Psi(\alpha)}=\cos{\alpha}\ket{00}+\sin{\alpha}\ket{11}$,
	the concurrence of entanglement reads $C=|\sin{2\alpha}|$~\cite{PhysRevLett.80.2245}. Concretely, the maximal value of uncertainty function $f(\alpha)$ is related to the concurrence $C$ as $f(C)=(\sqrt{(1+C)(3+C)^2})/(32\sqrt{2})$, that is, the maximal value of the uncertainty functions to witness entanglement.
	Similarly, the details about other uncertainty functions $g$, $h$, and $k$ can be found in S3-A of SM~\cite{urSM}.
	Moreover, we study the quantum upper bounds of uncertainty functions under the circumstance of the generalized Werner state $\rho(\alpha, \eta)=\eta\,\ket{\Psi(\alpha)}\bra{\Psi(\alpha)}+\big[(1-\eta)/4\big]\openone\otimes\openone$, with $\eta\in[0,1]$ expressing the proportion of $\ket{\Psi(\alpha)}$ in the mixed state. For the inequality~(\ref{eq:ent-0}), the quantum bound reads $f(C,\eta)=\big(\sqrt{1+\eta C}\,[2+\eta(1+C)]\big)/(32\sqrt{2})$, which means the witness entanglement still holds in the case of mixed state. See more details in S3-B of SM~\cite{urSM}.
	
	\begin{figure}[t]
		\includegraphics[width=0.48\textwidth]{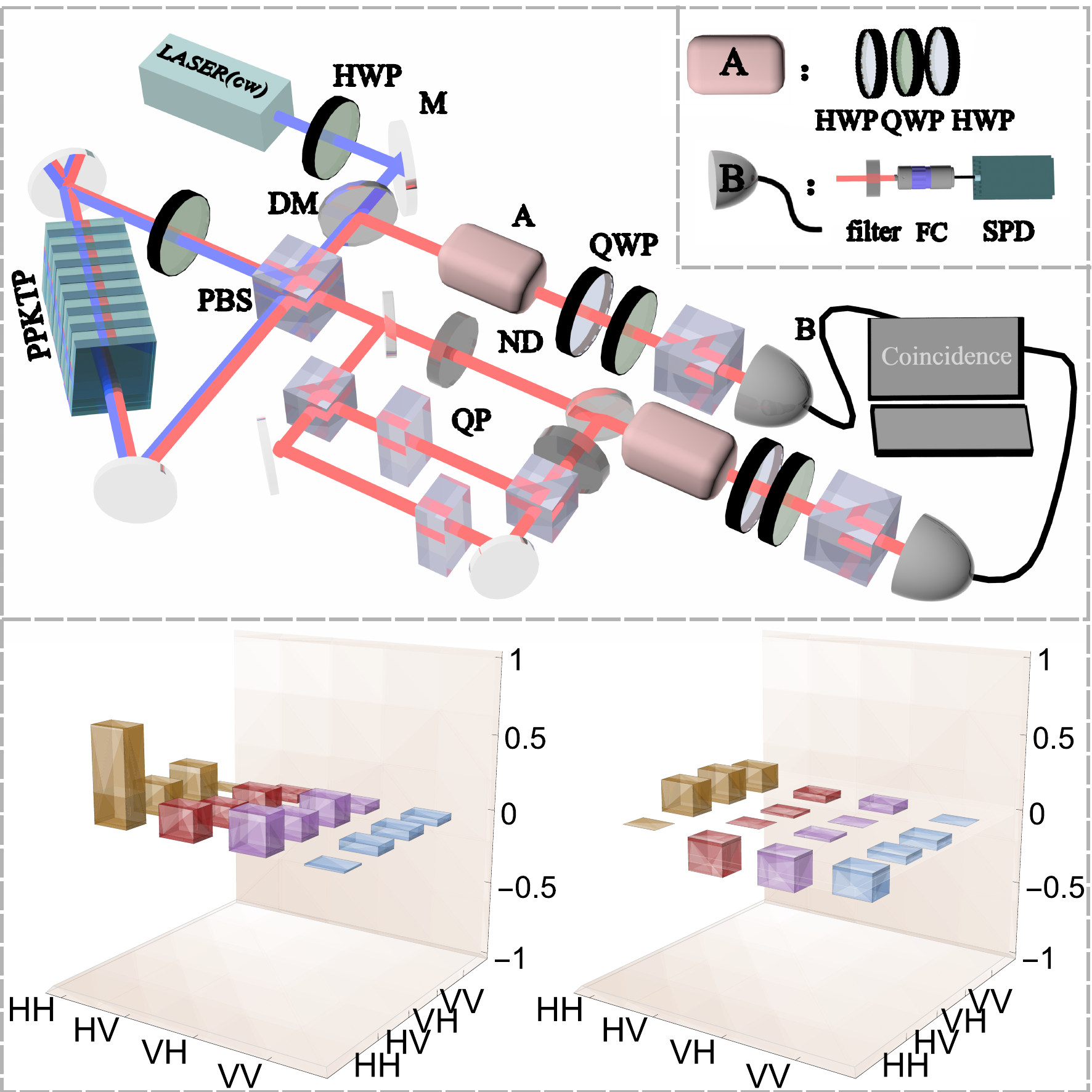}
		\caption{\textbf{The experiment setup and the experimental density matrix of $\rho_{12}$.} cw: continuous wave, HWP: half-wave plate, M: mirror, DM: dichroic mirror, PBS: polarized beam splitter, PPKTP: periodically poled KTiOPO4, QWP: quarter-wave plate, ND: neutral density filter, QP: quartz plate, FC: fiber coupler, SPD: single-photon detector. Section A is the ``optimizing box'' consisting of two HWPs and a QWP. Section B is the final setup including a filter, an FC, and a SPD. More details of the experiment setup are in the main text. The following two density matrices of the separable quantum state $\rho_{12}$ are the real (left) and imaginary (right) parts of the reconstructed experimental density matrix.
		}\label{expe}
	\end{figure}
	
	\emph{Experimental setup.} The experiment schematic is displayed in Fig. \ref{expe}. To prepare the two-qubit photon pairs in the Bell-type state $\ket{\Psi(\alpha)}=\cos\alpha\ket{\rm HH}+\sin\alpha\ket{\rm VV}$, we send a continuous-wave laser with a central wavelength of 404 nm to a piece of 20 mm-long periodically-poled KTiOPO$_4$ (PPKTP) nonlinear crystal clockwise and anticlockwise, where the induced type-II spontaneous parametric down-conversion process creates the degenerate photon pairs at 808 nm \cite{PhysRevLett.130.200202,Fedrizzi:07}.
	The parameter $\alpha$ is adjusted by the half-wave plate (HWP) after the pulse laser.
	Here, $\ket{\rm H}$ and $\ket{\rm V}$ respectively represent horizontal and vertical polarizations.

	We first certify the tight and non-trivial triple uncertainty relations in the two-qubit quantum systems through the separable quantum state $\rho_{12}=\ket{\psi_1}\otimes\ket{\psi_2}\bra{\psi_2}\otimes\bra{\psi_1}$, where $\ket{\psi_1}=\ket{\psi_2}=\sqrt{\frac{\sqrt{3}+1}{2\sqrt{3}}}\ket{\rm H}+\frac{1+\rm i}{\sqrt{2(3+\sqrt{3})}}\ket{\rm V}$.
	The quantum state $\rho_{12}$ not only satisfies the equal condition of Eq. (1), but also ensures the non-zero values on both sides.
	Moreover, due to the purity of $\rho_{12}$, it could be converted into the preparation of two single-qubit states as $\ket{\psi_1}$ and $\ket{\psi_2}$.
	Starting from $\ket{\Psi(\alpha=0)}=\ket{\rm HH}$, the two photons at $\ket{\rm H}$ are both directed to the combinations of an HWP, a quarter-wave plate (QWP) and an HWP (section A in Fig. \ref{expe}) to obtain $\ket{\psi_1}$ and $\ket{\psi_2}$ with parameters of wave plates determined through minimizing the trace distance between the parametric matrices and the single-qubit states ($\ket{\psi_1}$ and $\ket{\psi_2}$).
	
	To further prepare generalized Werner states $\rho(\alpha, \eta)$, one of the two photons is sent to the unbalanced interferometers (UIs) and the other one remains unchanged (see Fig. \ref{expe}).
	The UI separates the photons into three paths through a beam splitter (BS) and a PBS. 
	The transmitting path after the BS remains unchanged and the other two paths are inserted into a quartz plate (QP) to completely decohere the photon's polarization \cite{PhysRevLett.118.140404,PhysRevLett.130.200202}.
	Another PBS and BS combine the three paths so that an arbitrary Werner state $\rho(\alpha, \eta)$ could be prepared, with the parameter $\eta$ adjusted by the two neutral density filters (NDs).
	
	Then the two photons are both sent to the ``optimizing box'' (section A in Fig. \ref*{expe}) composed of a polarization rotation including two QWPs and one HWP.
	The rotation is a unitary operation $U=U_1\otimes U_2$ on the two-photon state, where
	\begin{align}
		\setlength{\arraycolsep}{1pt}
		\begin{array}{lc}
			U_r=
			\begin{pmatrix}
				\cos\theta_r & \sin\theta_r e^{-\rm i\phi_r}\\
				-\sin\theta_r e^{\rm i\phi_r} & \cos\theta_r
			\end{pmatrix},
		\end{array}
		\label{u}
	\end{align}
	$r=1,2$.
	We adjust the values of parameters $\theta_r$ and $\phi_r$ in operator $U_r$ to search for the maximal values of the uncertainty functions in \eqref{eq:ent-0}$\sim$\eqref{eq:ent-1}. 
	\begin{figure}[t]
		\includegraphics[width=0.49\textwidth]{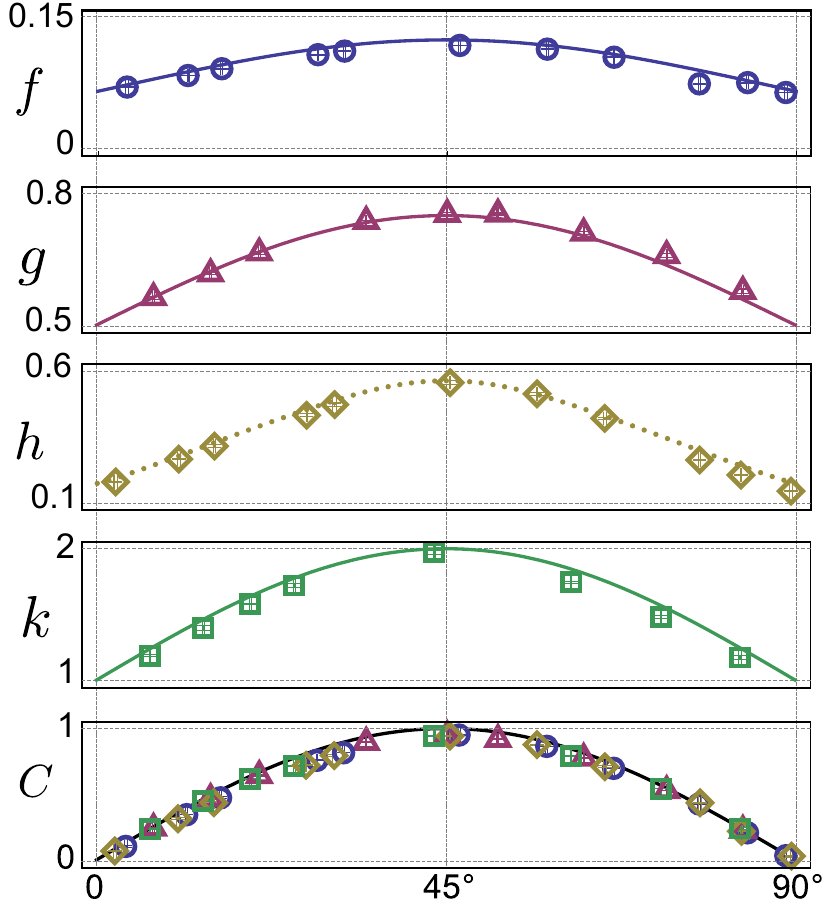}
		\caption{\textbf{Experiment results of Bell-type states.} The maximal values of four uncertainty functions in Eq. \eqref{eq:ent-0}$\sim$\eqref{eq:ent-1} and the degree of entanglement ($C$) versus the parameters $\alpha$ in $\ket{\Psi(\alpha)}$. The experimental results are in different types of hollow markers along with the theoretical analysis with solid ($f$, $g$, $k$) and dotted ($h$) curves. The error bars obtained through the Monte Carlo method are also shown in the figure. 
		}\label{fig1}
	\end{figure}
	To exemplify the optimizing procedure, we take $\alpha=0$ and $\alpha=45^\circ$ as examples, i.e., the state is $\ket{\Psi(\alpha=0)}=\ket{\rm HH}$ and the maximal value of $f$ is $3\sqrt{2}/64$ with the parameters of $U_1$ and $U_2$ are $\theta_1=-65.9^\circ$, $\theta_2=24.1^\circ$, $\phi_1=\phi_2=0$. And the state at $\ket{\Psi(\alpha=45^\circ)}=(1/\sqrt{2})(\ket{\rm HH}+\ket{\rm VV})$ corresponds to $f=1/8$ with $\theta_1=90^\circ$, $\theta_2=\phi_1=\phi_2=0$.
	The optimizing procedures are applied to search for the maximal values of uncertainty functions with the parameters in the operator $U$ varying with $\alpha$ in the quantum states, meanwhile satisfying the tight triple uncertainty relations for all quantum states. 
	The measurements of three components $\vec{J}_{i}$ along with their squares and the $\vec{R}_{i}$ ($i=x,y,z$) are implemented through the final QWP, HWP, and PBS to realize different joint polarization-projective measurements on the two outputs. 
	The section B in Fig. \ref{expe} with an interference filter with 3 nm bandwidth, a fiber coupler, and a single-photon detector for photon detection.
	The coincidence device is not shown in Fig. \ref{expe}. 
	Additionally, the final operators $J_i$, $R_i$ and $K_i$ are transformed into $\widetilde{J_i}$, $\widetilde{R_i}$ and $\widetilde{K_i}$ ($i=x,y,z$) in Eq. \eqref{eq:ent-0}$\sim$\eqref{eq:ent-1} through different unitary transformations. 
	
	\begin{figure}[t]
		\includegraphics[width=0.49\textwidth]{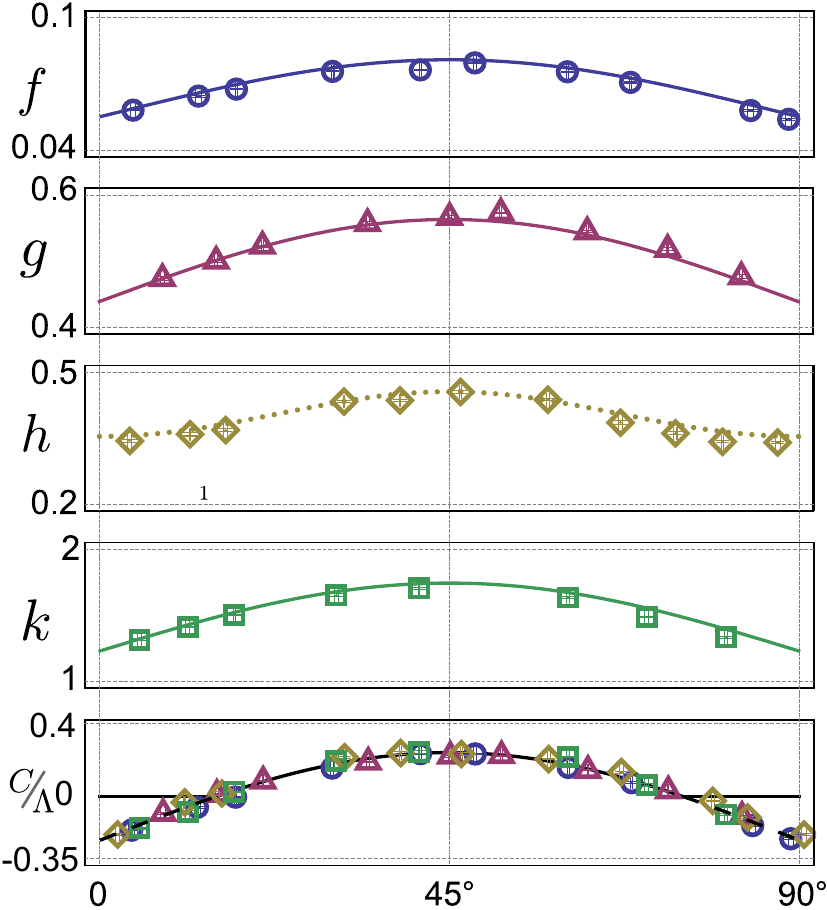}
		\caption{\textbf{Experiment results of the generalized Werner states.}
			The maximal values of the uncertainty functions in Eq. \eqref{eq:ent-0}$\sim$\eqref{eq:ent-1} and the degree of entanglement ($C$) versus the parameters $\alpha$ of the states $\rho(\alpha,\eta=0.5)$ ($\eta=0.5$ means mixed states). All experimental results are in different types of hollow markers along with the corresponding theoretical solid (in $f$, $g$ and $k$) and dotted curve (in $h$). In the last panel, the concurrence of the quantum states is in the same type of markers with the theoretical solid curves of the concurrence ($C$) and the dashed part representing the negative part of $\Lambda$. All error bars are calculated through the Monte Carlo method. 
		}\label{fig2}
	\end{figure}
	
	\emph{Results.}
	The quantum state $\rho_{12}$ is prepared with the fidelity of $F(\rho_{th},\rho_{exp})=\big(\mathrm{Tr}\sqrt{\sqrt{\rho_{th}}\rho_{exp}\sqrt{\rho_{th}}}\big)^2=99.2(1)\%$, where the $\rho_{th}$ and $\rho_{exp}$ respectively correspond to the theoretical and experimental density matrices. 
	The experimental density matrix $\rho_{exp}$ is also shown in Fig. \ref{expe}. 
	The obtained state is measured with the left term $\Delta J_x\Delta J_y\Delta J_z=0.026\pm0.005$, and the right term $|(\lambda^3/8)\langle R_x \rangle \langle R_y \rangle \langle R_z\rangle|^{1/2}=0.025\pm0.006$.
	The overlap between the two terms demonstrates the tight triple uncertainty relation is non-trivial under the constant $\lambda=2/\sqrt{3}$, which is also applicable to the additive form thereof, and the case of the total angular momentum $\vec{K}$ from both the product and additive occasions (the calculation details are listed in S1 and S2 of SM~\cite{urSM}). 
	Compared with the same tight constant in the one-qubit quantum system (spin-1/2) \cite{2014PRAKechrimparis,2017PRLfei}, the same tight constant $2/\sqrt{3}$ exploits its degree of universality and deepens the understanding of quantum uncertainty in formulating the significance of complex numbers in quantum mechanics \cite{RevModPhys.86.1261,fei2024}.
	
	We experimentally prepare a series of Bell-type states $\ket{\Psi(\alpha)}$ with the parameter $\alpha\in[0,90^\circ]$ and measure the maximal values of the uncertainty function in Eq. \eqref{eq:ent-0}$\sim$\eqref{eq:ent-1} corresponding to the product and additive forms of the tight triple uncertainty relations in the components of $\vec{J}$ and $\vec{K}$. 
	The unitary transformations in the ``the optimizing box'' (section A in Fig. \ref{expe}) are varied as the parameters $\alpha$ to obtain the maximum of four uncertainty functions. 
	In Fig. \ref{fig1}, the different types of hollow markers in the upper four panels indicate the experiment-measured results of the uncertainty functions ranging from $f$, $g$, $h$ to $k$. 
	The last panel shows the relation between the degree of entanglement (through concurrence $C$) and the angle parameter ($\alpha$). The concurrence $C=\max\{0,\Lambda=\lambda_1-\lambda_2-\lambda_3-\lambda_4\}$ is calculated through the experimentally reconstructed density matrix, with $\lambda_i$ corresponding to the square root of the decreasing eigenvalues of Hermitian matrix $\rho(\sigma_y\otimes\sigma_y)\rho^*(\sigma_y\otimes\sigma_y)$, $\sigma_y$ representing the $y$-component of the Pauli operator, and $\rho^*$ denoting the complex conjugate of $\rho$ \cite{PhysRevLett.80.2245}. 
	Our experiment results are in good accordance with the corresponding theoretical results and demonstrate that the maximal values of uncertainty functions are related to the degree of entanglement. 
	Here, the maximal values of uncertainty functions $f$, $g$ and $k$ have exact analytical solutions corresponding to the solid curves, however, $h$ is represented with the exact theoretical analytic dots (without analytical solution) in Fig. \ref{fig1}. 
	See more theoretical analysis of the maximal values of uncertainty functions in the \cite{urSM}. 
	We also note that the higher degree of entanglement directly correlates with the increased maximum of uncertainty functions, and the distinct undulations of the uncertainty functions would exhibit a higher degree of sensitivity in quantifying concurrence. 
	The above conclusions are also satisfied when parameters $\alpha$ are outside $[0,90^\circ]$, that is, the four uncertainty functions could be regarded as periodic functions with a period $\pi/2$. 
	
	To ensure the generality, the Werner states $\rho(\alpha,\eta)$ are further prepared and explored with the maximal values of the uncertainty functions in Eq. \eqref{eq:ent-0}$\sim$\eqref{eq:ent-1}.
	Fig. \ref{fig2} shows the experimental results of the maximal values of four uncertainty functions varying with the $\alpha\in[0,90^\circ]$ and $\eta=0.5$, along with the corresponding theoretical analysis results (solid curves in $f$, $g$ and $k$ and dotted curve in $h$). 
	The experimental concurrence is also exhibited in the same type markers, with the theoretical solid and dashed parts representing $\Lambda\le0$. 
	We give more theoretical analysis in the SM~\cite{urSM}. 
	Our experimental results are in good agreement with the theoretical results, thereby substantiating the viability and robustness of our uncertainty functions even when applied to quantum states with noise. 
	Compared with the pure cases ($\eta=1$), higher values of $\eta$ correspond to the increased maximum of uncertainty functions at the point $\alpha=45^\circ$. When $\eta=0$, the quantum states $\rho(\alpha, \eta)$ are totally mixed and the four uncertainty functions are not varied with the parameter $\alpha$.

	\emph{Conclusions.}
	In this paper, we advance the study of the tight triple uncertainty relations based on the two-qubit quantum systems with experimental demonstrations in versatile optical setups and reveal the generality of the tight constant $2/\sqrt{3}$. 
	We experimentally verify the attainable non-trivial bound of the uncertainty relations, and demonstrate the maximal value of the uncertainty functions is associated with the concurrence of entanglement through a series of pure and mixed quantum states. 
	Particularly, the extensions to generalized Werner states with white noise would simulate more potentials in the open quantum scenarios \cite{PhysRevLett.126.010602}.
	
	Our work would lead to more innovative applications of the uncertainty principle in quantum metrology and quantum information science, e.g., the uncertainty relations in the non-Hermitian quantum systems \cite{PhysRevLett.132.070203}. Furthermore, this may deepen the understanding of the inner associations between the measurement incompatibility and quantum nonlocality.

	This work was supported by the Innovation Program for Quantum Science and Technology (Grants No. 2021ZD0301200 and No. 2021ZD0301400), the National Natural Science Foundation of China (Grants No. 61975195, No. 11821404, No. 92365205, No. 12374336, No. 11874343, No. 62475249 and No. 12404403), the Anhui Initiative in Quantum Information Technologies (Grant No. AHY060300), USTC Research Funds of the Double First-Class Initiative (Grant No. YD2030002024) and USTC Major Frontier Research Program LS2030000002. 
	J.L.C. is supported by the National Natural Science Foundation of China (Grants No. 12275136 and 12075001) and the 111 Project of B23045.

	Y.W. and J.Z. contributed equally to this work.
	
	\section{The new triple uncertainty relation for $\vec{J}$ in 2-qubit system}
	
	For any two observables $A$ and $B$, the Heisenberg-Robertson uncertainty relation~\cite{1929PRRobertson}
	\begin{align}\label{eq:1}
		\Delta A \Delta B\geq \frac{1}{2}|\langle[A,B]\rangle|,
	\end{align}
	with the standard deviation of the observable $A, B$ is
	\begin{align}\label{eq:2}
		\Delta A=\sqrt{\langle A^2\rangle-\langle A\rangle^2},~~~\Delta B=\sqrt{\langle B^2\rangle-\langle B\rangle^2},
	\end{align}
	the angle brackets $\langle \rangle$ denoting the expectation of an operator with respect to a given state $\rho$, and the commutation relation of operators $A$ and $B$ is $[A, B]=A B-B A$.
	
	We suppose the operator $\vec{J}$ as
	\begin{align}\label{eq:2ur-1}
		\vec{J}=\vec{S}_1 \times \vec{S}_2,
	\end{align}
	where the angular momentum for the $i$th qubit as $S_i=(S_{ix}, S_{iy}, S_{iz})$.
	
	Thus, the three components of operator $J$ are
	\begin{align}\label{eq:2ur-2}
		J_x&=S_{1y}S_{2z}-S_{1z}S_{2y},\\
		J_y&=S_{1z}S_{2x}-S_{1x}S_{2z},\\
		J_z&=S_{1x}S_{2y}-S_{1y}S_{2x}.
	\end{align}
	
	By calculation, we have the commutation relations of $J_i$ as
	\begin{align}\label{eq:2ur-3}
		[J_x, J_y]&=\frac{1}{4}i(S_{1z}+S_{2z}),\\
		[J_y, J_z]&=\frac{1}{4}i(S_{1x}+S_{2x}),\\
		[J_z, J_x]&=\frac{1}{4}i(S_{1y}+S_{2y}).
	\end{align}
	
	Let's define
	\begin{align}\label{eq:2ur-4}
		R_x&\equiv\frac{1}{4}(S_{1x}+S_{2x}),\\
		R_y&\equiv\frac{1}{4}(S_{1y}+S_{2y}),\\
		R_z&\equiv\frac{1}{4}(S_{1z}+S_{2z}).
	\end{align}
	
	According to Eq.~(\ref{eq:1}), the uncertainty relation between each two components of $J=(J_x, J_y, J_z)$ as
	\begin{align}\label{eq:2ur-5}
		\Delta J_x \Delta J_y &\geq \frac{1}{2}|\langle[J_x,J_y]\rangle|=\frac{1}{2}|\langle R_z\rangle|,\nonumber\\
		\Delta J_y \Delta J_z &\geq \frac{1}{2}|\langle[J_y,J_z]\rangle|=\frac{1}{2}|\langle R_x\rangle|,\nonumber\\
		\Delta J_z \Delta J_x &\geq \frac{1}{2}|\langle[J_z,J_x]\rangle|=\frac{1}{2}|\langle R_y\rangle|.
	\end{align}

	\subsection{The product form uncertainty relation of $\vec{J}$}

	By multiplying the above uncertainty relations (\ref{eq:2ur-5}) and then taking the square root, a trivial uncertainty relation can be obtained
	\begin{align}\label{eq:2ur-6}
		\Delta J_x \Delta J_y \Delta J_z &\geq |\frac{1}{8}\langle R_x\rangle \langle R_y\rangle \langle R_z\rangle|^{1/2},
	\end{align}
	it is not tight, there is no state satisfying such a lower bound. Below, we study the tightness of the uncertainty relation.
	
	We introduce the triple constant $\lambda$ and suppose the tight uncertainty relation as
	\begin{align}\label{eq:2ur-7}
		\Delta J_x \Delta J_y \Delta J_z &\geq |\frac{\lambda^3}{8}\langle R_x\rangle \langle R_y\rangle \langle R_z\rangle|^{1/2}.
	\end{align}
	
	Below, we look for the value of $\lambda$ that satisfy the above tight inequality.
	
	We choose the arbitrary two-qubit state
	\begin{align}\label{eq:10}
		\rho_{AB}=\frac{1}{4}(I_2\otimes I_2+\Vec{a}\cdot\Vec{\sigma}\otimes I_2+I_2\otimes\Vec{b}\cdot\Vec{\sigma}+\sum_{ij}T_{ij}\sigma_i\otimes\sigma_j),
	\end{align}
	where the Bloch vector $\Vec{a}=(a_1,a_2,a_3), \Vec{b}=(b_1,b_2,b_3)$, pauli matrix $\Vec{\sigma}=(\sigma_x,\sigma_y,\sigma_z)$, and the spin correlation matrix
	\begin{align}
		T=\left(
		\begin{array}{ccc}
			T_{11} & T_{12} & T_{13}  \\
			T_{21} & T_{22} & T_{23}  \\
			T_{31} & T_{32} & T_{33}  \\
		\end{array}
		\right).
	\end{align}
	
	Taking Eq.~(\ref{eq:10}) into Eq.~(\ref{eq:2}), we can get the variances of $J_i$ as
	\begin{align}
		(\Delta J_x)^2&=\langle J_x^2\rangle-\langle J_x\rangle^2\nonumber\\
		&=\frac{2-2 T_{11}-T_{23}^2+2 T_{23}T_{32} -T_{32}^2}{16},\\
		(\Delta J_y)^2&=\langle J_y^2\rangle-\langle J_y\rangle^2\nonumber\\
		&=\frac{2-2 T_{22}-T_{13}^2+2 T_{13}T_{31} -T_{31}^2}{16},\\
		(\Delta J_z)^2&=\langle J_z^2\rangle-\langle J_z\rangle^2\nonumber\\
		&=\frac{2-2 T_{33}-T_{12}^2+2 T_{12}T_{21} -T_{21}^2}{16},
	\end{align}
	and through calculation, one obtains
	\begin{align}
		|\langle R_x\rangle|&=|\frac{a_1+b_1}{8}|,\\
		|\langle R_y\rangle|&=|\frac{a_2+b_2}{8}|,\\
		|\langle R_z\rangle|&=|\frac{a_3+b_3}{8}|.
	\end{align}
	
	According to the symmetry, we have
	\begin{align}\label{eq:SymmetryAB}
		a1=a2=a3=a,~~~b1=b2=b3=b,
	\end{align}
	and
	\begin{align}\label{eq:SymmetryT}
		&T_{11}=T_{22}=T_{33}=t_1,~~~T_{12}=T_{21}=t_2,\nonumber\\
		&T_{13}=T_{31}=t_3,~~~~~~~~~~~T_{23}=T_{32}=t_4.
	\end{align}
	
	We can rewrite the tight uncertainty relation as
	\begin{align}\label{eq:tau1}
		\lambda(t_1, t_2, t_3, t_4) \leq (\frac{64 (\Delta J_x)^4 (\Delta J_y)^4 (\Delta J_z)^4}{(\langle R_x\rangle)^2 (\langle R_y\rangle)^2 (\langle R_z\rangle)^2})^{\frac{1}{6}}.
	\end{align}
	
	\begin{figure}[htbp]
		\centering
		\includegraphics[width=8cm]{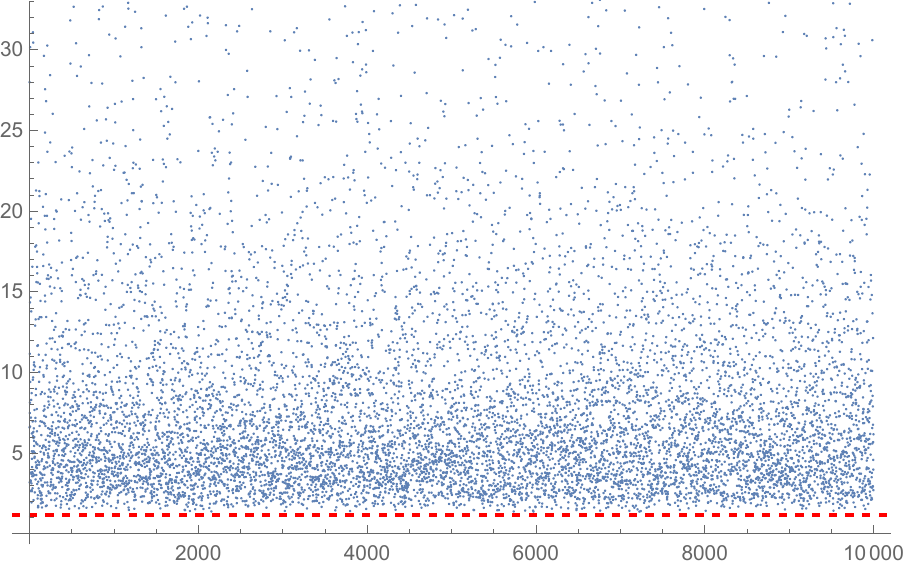}
		\caption{The scatter diagram of $\lambda(t_1, t_2, t_3, t_4)$ of Eq.~(\ref{eq:tau1}), the red line is $2/\sqrt{3}$.}
	\end{figure}

	By going through all of $t_1, t_2, t_3, t_4$, we find
	\begin{align}
		\min[(\frac{64 (\Delta J_x)^4 (\Delta J_y)^4 (\Delta J_z)^4}{(\langle R_x\rangle)^2 (\langle R_y\rangle)^2 (\langle R_z\rangle)^2})^{\frac{1}{6}}]=\frac{2}{\sqrt{3}},
	\end{align}
	then
	\begin{align}
		\lambda(t_1, t_2, t_3, t_4) \leq \frac{2}{\sqrt{3}},
	\end{align}
	i.e., we can obtain the triple constant
	\begin{align}
		\lambda=\frac{2}{\sqrt{3}}.
	\end{align}

	\emph{Remark 1.---}The equal sign of trivial uncertainty relation (\ref{eq:2ur-7}) holds when the two-qubit state is a separable state $\ket{\psi_{AB}}$, which can transform to the state $\ket{00}$.
	
	In detail,
	\begin{align}\label{eq:MinCondition}
		a=b=\dfrac{1}{\sqrt{3}},~t_1=t_2=t_3=t_4=\frac{1}{3},
	\end{align}
	
	\begin{align}
		(\Delta J_x)^2=(\Delta J_y)^2=(\Delta J_z)^2=\frac{1}{12},
	\end{align}
	\begin{align}
		|\langle R_x\rangle|=|\langle R_y\rangle|=|\langle R_z\rangle|=\frac{1}{4\sqrt{3}}.
	\end{align}
	
	The two-qubit state $\rho_{12}$ is a separable state
	\begin{align}
		\rho_{12}=\ket{\psi_{12}}\bra{\psi_{12}},
	\end{align}
	with
	\begin{align}
		\ket{\psi_{12}}=\ket{\psi_{1}}\otimes\ket{\psi_{2}},
	\end{align}
	where
	\begin{align}
		\ket{\psi_{1}}=\sqrt{\frac{\sqrt{3}+1}{2\sqrt{3}}}\ket{0}+\frac{1+\rm i}{\sqrt{2(3+\sqrt{3})}}\ket{1},
	\end{align}
	and
	\begin{align}
		\ket{\psi_{2}}=-\rm i(\sqrt{\frac{\sqrt{3}+1}{2\sqrt{3}}}\ket{0}+\frac{1+\rm i}{\sqrt{2(3+\sqrt{3})}}\ket{1}),
	\end{align}
	with $\rm i^2=-1$.
	
	The separable state $\ket{\psi_{12}}$ can transform the state $\ket{00}$, through a unitary operator $U$,
	\begin{align}
		U=\rm i \left(
		\begin{array}{cccc}
			\frac{3+\sqrt{3}}{6} & \frac{1-\rm i}{2\sqrt{3}} & \frac{1-\rm i}{2\sqrt{3}} & \frac{\rm i(\sqrt{3}-3)}{6} \\
			-\sqrt{\frac{\rm i}{6}} & \frac{3+\sqrt{3}}{6} & -\sqrt{\frac{2-\sqrt{3}}{6}} & \frac{1-\rm i}{2\sqrt{3}} \\
			-\sqrt{\frac{\rm i}{6}} & -\sqrt{\frac{2-\sqrt{3}}{6}} & \frac{3+\sqrt{3}}{6} & \frac{1-\rm i}{2\sqrt{3}}
			\\
			-\frac{\rm i(\sqrt{3}-3)}{6} & -\sqrt{\frac{\rm i}{6}} & -\sqrt{\frac{\rm i}{6}} & \frac{3+\sqrt{3}}{6} \\
		\end{array}
		\right).
	\end{align}
	This means that the uncertainty relation has equal sign when the state is separable.

	\subsection{The additive form uncertainty relation of $\vec{J}$}

	According to $a^2+b^2\geq 2ab$ and Eq.~(\ref{eq:2ur-5}), we have
	\begin{align}
		(\Delta J_x)^2+(\Delta J_y)^2 &\geq |\langle R_z\rangle|,\\
		(\Delta J_y)^2+(\Delta J_z)^2 &\geq |\langle R_x\rangle|,\\
		(\Delta J_z)^2+(\Delta J_x)^2 &\geq |\langle R_y\rangle|,
	\end{align}
	then,
	\begin{align}
		(\Delta J_x)^2+(\Delta J_y)^2+(\Delta J_z)^2 &\geq \frac{1}{2}(|\langle R_x\rangle|+|\langle R_y\rangle|+|\langle R_z\rangle|),
	\end{align}
	we also tighten the lower bound by introducing the triple constant $\lambda$,
	\begin{align}\label{eq:Jadd}
		(\Delta J_x)^2+(\Delta J_y)^2+(\Delta J_z)^2 &\geq \frac{\lambda}{2}(|\langle R_x\rangle|+|\langle R_y\rangle|+|\langle R_z\rangle|).
	\end{align}
	
	Similarly,
	\begin{align}\label{eq:tau3}
		\lambda(t_1, t_2, t_3, t_4) \leq \frac{2 ((\Delta J_x)^2+(\Delta J_y)^2+(\Delta J_z)^2)}{|\langle R_x\rangle|+|\langle R_y\rangle|+ |\langle R_z\rangle|}).
	\end{align}
	
	For the arbitrary two-qubit state, by going through all of $t_1, t_2, t_3, t_3$, we find
	\begin{align}
		Min [\frac{2 ((\Delta J_x)^2+(\Delta J_y)^2+(\Delta J_z)^2)}{|\langle R_x\rangle|+|\langle R_y\rangle|+ |\langle R_z\rangle|})]=\frac{2}{\sqrt{3}},
	\end{align}
	i.e.,
	\begin{align}
		\lambda(t_1, t_2, t_3, t_4) \leq \frac{2}{\sqrt{3}},
	\end{align}
	so, we can obtain the triple constant
	\begin{align}
		\lambda=\frac{2}{\sqrt{3}}.
	\end{align}

	\begin{figure}[htbp]
		\centering
		\includegraphics[width=8cm]{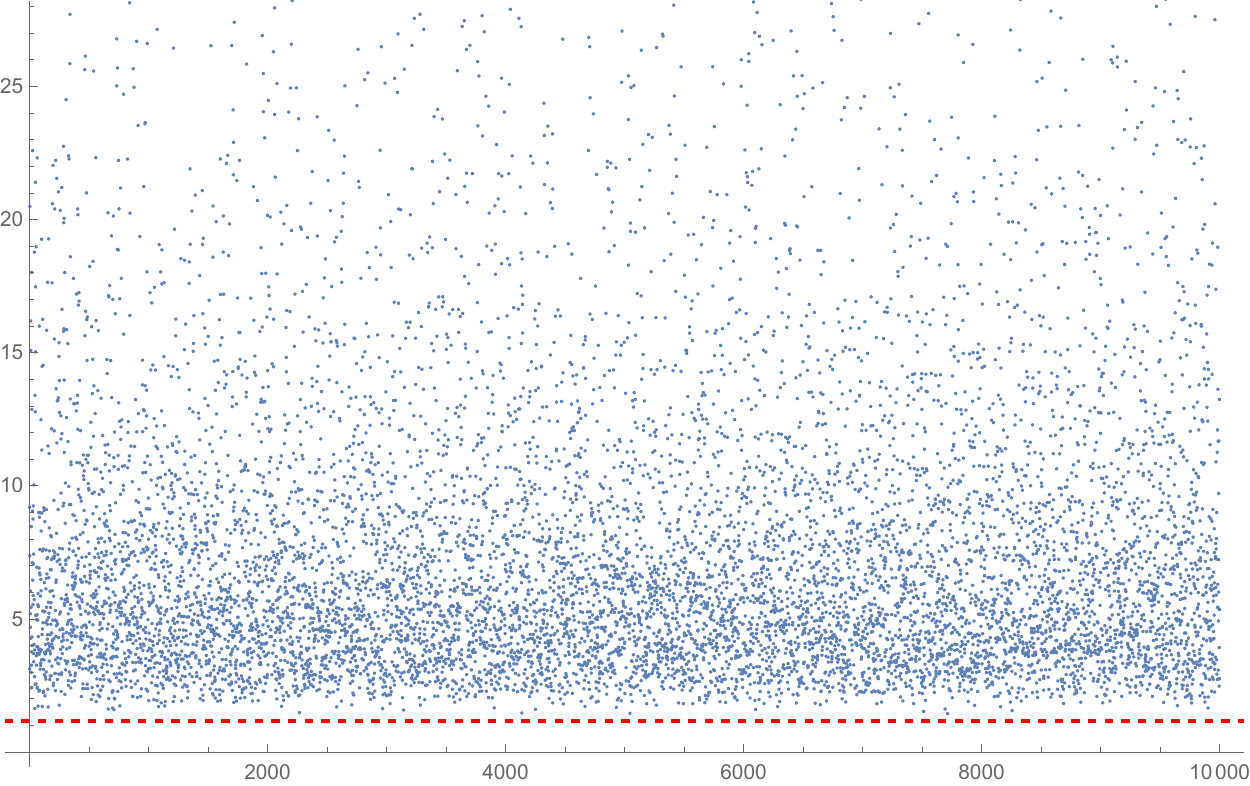}
		\caption{The scatter diagram of $\lambda(t_1, t_2, t_3, t_4)$ of Eq.~(\ref{eq:tau3}), the red line is $2/\sqrt{3}$.}
	\end{figure}

	\section{The new triple uncertainty relation for $\vec{K}$ in 2-qubit system}
	
	Define the total angular momentum for a two-qubit system  as
	\begin{equation}
		\vec{K}=\vec{S}_1\otimes\openone+\openone\otimes\vec{S}_2,
	\end{equation}
	whose components read
	\begin{equation}
		K_l=S_{1l}\otimes\openone+\openone\otimes S_{2l}=4\,R_l,\quad l=x,y,z.
	\end{equation}
	
	Then
	\begin{equation}\label{eq:KlSqure}
		\begin{split}
			K_l^2 &= \left(S_{1l}+S_{2l}\right)\left(S_{1l}+S_{2l}\right) \\
			&= \dfrac{\openone\otimes\openone}{2}+S_{1l}\,S_{2l}+S_{2l}\,S_{1l} \\
			&= \dfrac{\openone\otimes\openone+\,\sigma_{1l}\,\sigma_{2l}}{2},
		\end{split}
	\end{equation}
	which means that for any two-qubit state \eqref{eq:10},
	\begin{equation}\label{eq:ExvKlSqure}
		\begin{split}
			\exv{K_l^2}
			&=tr{[\rho_{AB} K_l^2]} \nonumber\\
			&=\frac{1+T_{ll}}{2},
		\end{split}
	\end{equation}
	and
	\begin{equation}
		\exv{K_l}=4\exv{R_l}=\frac{a_l+b_l}{2}.
	\end{equation}
	Therefore,
	\begin{equation}\label{eq:DeltaK}
		\begin{split}
			\Delta K_l &= \sqrt{\exv{K_l^2}-\exv{K_l}^2} \\
			&= \sqrt{\dfrac{1+T_{ll}}{2}-\left(\frac{a_l+b_l}{2}\right)^2} \\
			&= \dfrac{\sqrt{2(1+t_1)-(a+b)^2}}{2}.
		\end{split}
	\end{equation}
	Note the last equal sign in Eq.~\eqref{eq:DeltaK} holds under the assumption of symmetry \eqref{eq:SymmetryAB} and
	\eqref{eq:SymmetryT}. Moreover, the commutative relation of any two components of the total angular momentum operator $\vec{K}$ is
	of the following form
	\begin{equation}
		\begin{split}
			\left[K_l,K_m\right] &= \left[S_{1l}+S_{2l},\,S_{1m}+S_{2m}\right] \\
			&= \left[S_{1l},S_{1m}\right]+\left[S_{2l},S_{2m}\right] \\
			&= \varepsilon_{lmn}\left(S_{1n}+S_{2n}\right) \\
			&= \varepsilon_{lmn}\,K_n,
		\end{split}
	\end{equation}
	where $l,m,n=x,y,z$, and $\varepsilon_{lmn}$ represents the Levi-Civita symbol.

	\subsection{The product form uncertainty relation of $\vec{K}$}

	By dint of Eq.~(\ref{eq:1}), the uncertainty relation between two components of $\vec{K}$ can be expressed as
	\begin{equation}\label{eq:dkdk}
		\Delta K_l\,\Delta K_m\geq\frac{1}{2}\Bigl\lvert\bigl\langle\left[K_l,K_m\right]\bigr\rangle\Bigr\lvert
		=\frac{1}{2}\bigl|\langle K_n\rangle\bigr|.
	\end{equation}
	By the same token, we can deduce the uncertainty relation using the three components of $\vec{K}$ as follows
	\begin{align}\label{eq:KMultiple}
		\Delta K_x\,\Delta K_y\,\Delta K_z\geq\left|\frac{1}{8}\exv{K_x}\exv{K_y}\exv{K_z}\right|^{1/2}.
	\end{align}
	To tighten the inequality \eqref{eq:KMultiple}, we cite the triple constant $\lambda$, i.e.
	\begin{align}\label{eq:KMultipleTight}
		\Delta K_x\,\Delta K_y\,\Delta K_z\geq\left|\frac{\lambda^3}{8}\exv{K_x}\exv{K_y}\exv{K_z}\right|^{1/2}.
	\end{align}
	After minimizing the function
	\begin{equation}\label{eq:tau2}
		\lambda(a,b,t_1,t_2,t_3,t_4)\leq(\frac{64(\Delta K_x)^4(\Delta K_y)^4(\Delta K_z)^4}{(\exv{K_x})^2(\exv{K_y})^2(\exv{K_z})^2})^\frac{1}{6},
	\end{equation}
	under legal $\rho_{AB}$, we attain the critical $\lambda=2/\sqrt{3}$. The minimum is saturated under the condition
	of \eqref{eq:MinCondition}.
	
	\begin{figure}[htbp]
		\centering
		\includegraphics[width=8cm]{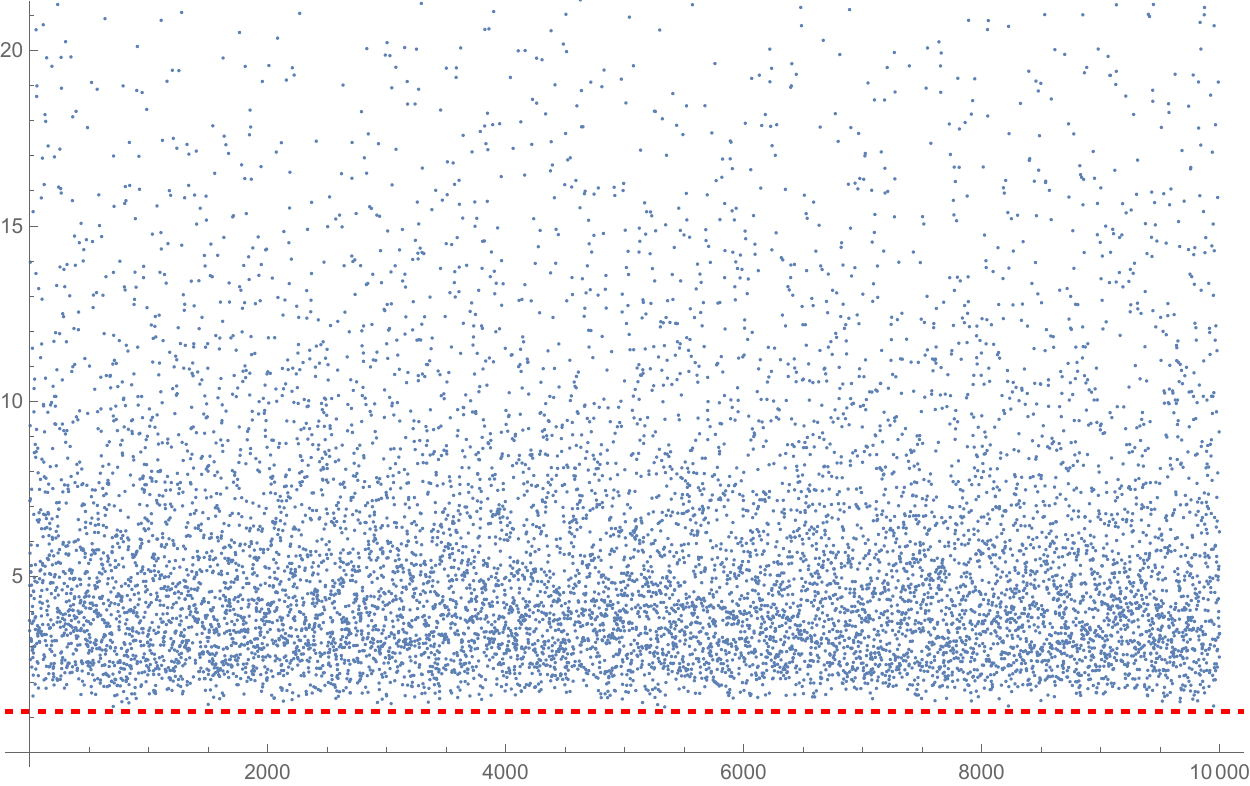}
		\caption{The scatter diagram of $\lambda(t_1, t_2, t_3, t_4)$ of (\ref{eq:tau2}), the red line is $2/\sqrt{3}$.}
	\end{figure}

	\subsection{The additive form uncertainty relation of $\vec{K}$}
	
	Similarly, according to $a^2+b^2\geq 2ab$ and Eq.~(\ref{eq:dkdk}), we have
	\begin{align}
		(\Delta K_x)^2+(\Delta K_y)^2 &\geq |\langle K_z\rangle|,\\
		(\Delta K_y)^2+(\Delta K_z)^2 &\geq |\langle K_x\rangle|,\\
		(\Delta K_z)^2+(\Delta K_x)^2 &\geq |\langle K_y\rangle|,
	\end{align}
	then,
	\begin{align}
		(\Delta K_x)^2+(\Delta K_y)^2+(\Delta K_z)^2 &\geq \frac{1}{2}(|\langle K_x\rangle|+|\langle K_y\rangle|+|\langle K_z\rangle|),
	\end{align}
	the tight uncertainty in addition version reads
	\begin{align}\label{eq:Kadd}
		(\Delta K_x)^2+(\Delta K_y)^2+(\Delta K_z)^2 &\geq \frac{\lambda}{2}(|\langle K_x\rangle|+|\langle K_y\rangle|+|\langle K_z\rangle|).
	\end{align}
	
	Similarly,
	\begin{align}\label{eq:tau4}
		\lambda(t_1, t_2, t_3, t_4) \leq \frac{2 ((\Delta K_x)^2+(\Delta K_y)^2+(\Delta K_z)^2)}{|\langle K_x\rangle|+|\langle K_y\rangle|+ |\langle K_z\rangle|}).
	\end{align}
	
	For the arbitrary two-qubit state, by going through all of $t_1, t_2, t_3, t_3$, we find
	\begin{align}
		Min [\frac{2 ((\Delta K_x)^2+(\Delta K_y)^2+(\Delta K_z)^2)}{|\langle K_x\rangle|+|\langle K_y\rangle|+ |\langle K_z\rangle|})]=\frac{2}{\sqrt{3}},
	\end{align}
	i.e.,
	\begin{align}
		\lambda(t_1, t_2, t_3, t_4) \leq \frac{2}{\sqrt{3}},
	\end{align}
	so, we can obtain the triple constant
	\begin{align}
		\lambda=\frac{2}{\sqrt{3}}.
	\end{align}

	\begin{figure}[htbp]
		\centering
		\includegraphics[width=8cm]{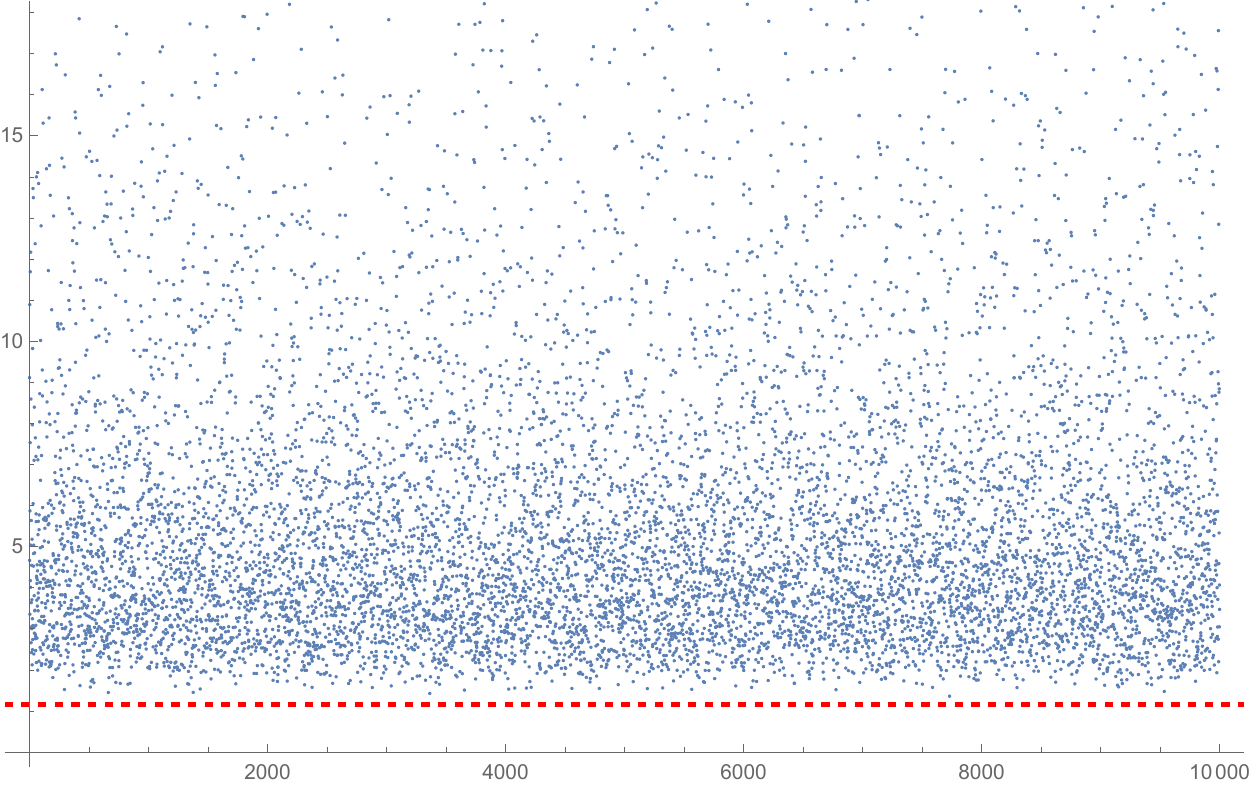}
		\caption{The scatter diagram of $\lambda(t_1, t_2, t_3, t_4)$ of Eq.~(\ref{eq:tau4}), the red line is $2/\sqrt{3}$.}
	\end{figure}

	\section{Four uncertainty functions $f, h, g$, and $k$}
	
	Four uncertainty functions can be constructed from uncertainty relation (\ref{eq:2ur-7}), (\ref{eq:Jadd}), (\ref{eq:KMultipleTight}) and (\ref{eq:Kadd}),
	\begin{align}\label{eq:bellJpro}
		f=\Delta \widetilde{J_x} \Delta \widetilde{J_y} \Delta \widetilde{J_z}
		- \left\lvert\frac{\lambda^3}{8}\langle\widetilde{R_x}\rangle \langle \widetilde{R_y}\rangle \langle \widetilde{R_z}\rangle\right\lvert^{1/2}\geq 0,
	\end{align}
	
	\begin{equation}\label{eq:bellJadd}
		g=\sum_k \left(\Delta\widetilde{J_k}\right)^2-\frac{1}{\sqrt{3}}\left(\sum_k\abs{ \exv{\widetilde{R_k}}}\right)\geq 0,
	\end{equation}
	
	\begin{align}\label{eq:bellKpro}
		h=\Delta \widetilde{K_x} \Delta \widetilde{K_y} \Delta \widetilde{K_z}
		- \left\lvert\frac{\lambda^3}{8}\langle\widetilde{K_x}\rangle \langle \widetilde{K_y}\rangle \langle \widetilde{K_z}\rangle\right\lvert^{1/2}\geq 0,
	\end{align}
	
	\begin{equation}\label{eq:bellKadd}
		k=\sum_l \left(\Delta\widetilde{K_l}\right)^2-\frac{1}{\sqrt{3}}\left(\sum_l\abs{ \exv{\widetilde{K_l}}}\right)\geq 0,
	\end{equation}
	with $\lambda=2/\sqrt{3}$, the operator
	\begin{align}
		\widetilde{O_i}=UO_iU^\dagger,
	\end{align}
	where $O=J, K, R$, and $i=x, y, z$, the unitary operator
	\begin{align}
		U=U_1\otimes U_2,
	\end{align}
	where
	\begin{align}\label{eq:ent-3}
		U_i=\left(
		\begin{array}{cc}
			\cos{\theta_r} & \sin{\theta_r e^{-\rm i\phi_r}} \\
			-\sin{\theta_r e^{\rm i\phi_r}} & \cos{\theta_r} \\
		\end{array}
		\right),
	\end{align}
	$r=1, 2$.

	\subsection{The maximal value of four uncertainty functions with any 2-qubit pure state}
	
	Any 2-qubit pure state as
	\begin{align}\label{eq:ent-5}
		\ket{\Psi(\alpha)}=\cos{\alpha}\ket{00}+\sin{\alpha}\ket{11},
	\end{align}
	the concurrence of entanglement reads~\cite{PhysRevLett.80.2245}
	\begin{align}
		C=|\sin{2\alpha}|.
	\end{align}
	
	In the following, we study the maximal quantum values of four uncertainty functions $f, h, g$, and $k$ under 2-qubit pure state.

	\begin{enumerate}
		\item For uncertainty function $f$, the maximal quantum value reads
		\begin{align}\label{eq:ent-6}
			f(\alpha)=\frac{\sqrt{1+\sin{2\alpha}}(3+\sin{2\alpha})}{32\sqrt{2}},
		\end{align}
		where $\alpha\in[0,\pi/2]$.
		Then, the quantum violation $f(\alpha)$ is related to the concurrence $C$ as
		\begin{align}
			f(C)=\frac{\sqrt{(1+C)(3+C)^2}}{32\sqrt{2}},
		\end{align}
		that is, the maximal value of the uncertainty functions $f$ to witness entanglement.
		
		According to caculation and the Fig.~\ref{fig:fa}, when $\alpha=\pi/4$, i.e., the state is maximum entangled state
		$\ket{\Psi(\alpha=\pi/4)}=\frac{1}{\sqrt{2}}(\ket{00}+\ket{11})$, we can obtain the maximum quantum value $f(\frac{\pi}{4})=1/8$.
		At this point, the parameters of the unitary operator $U_i$ in Eq.~(\ref{eq:ent-3}) are $\theta_1=\pi/2, \theta_2=\phi_1=\phi_2=0$.
		
		\begin{figure}[h]
			\centering
			\includegraphics[width=8cm]{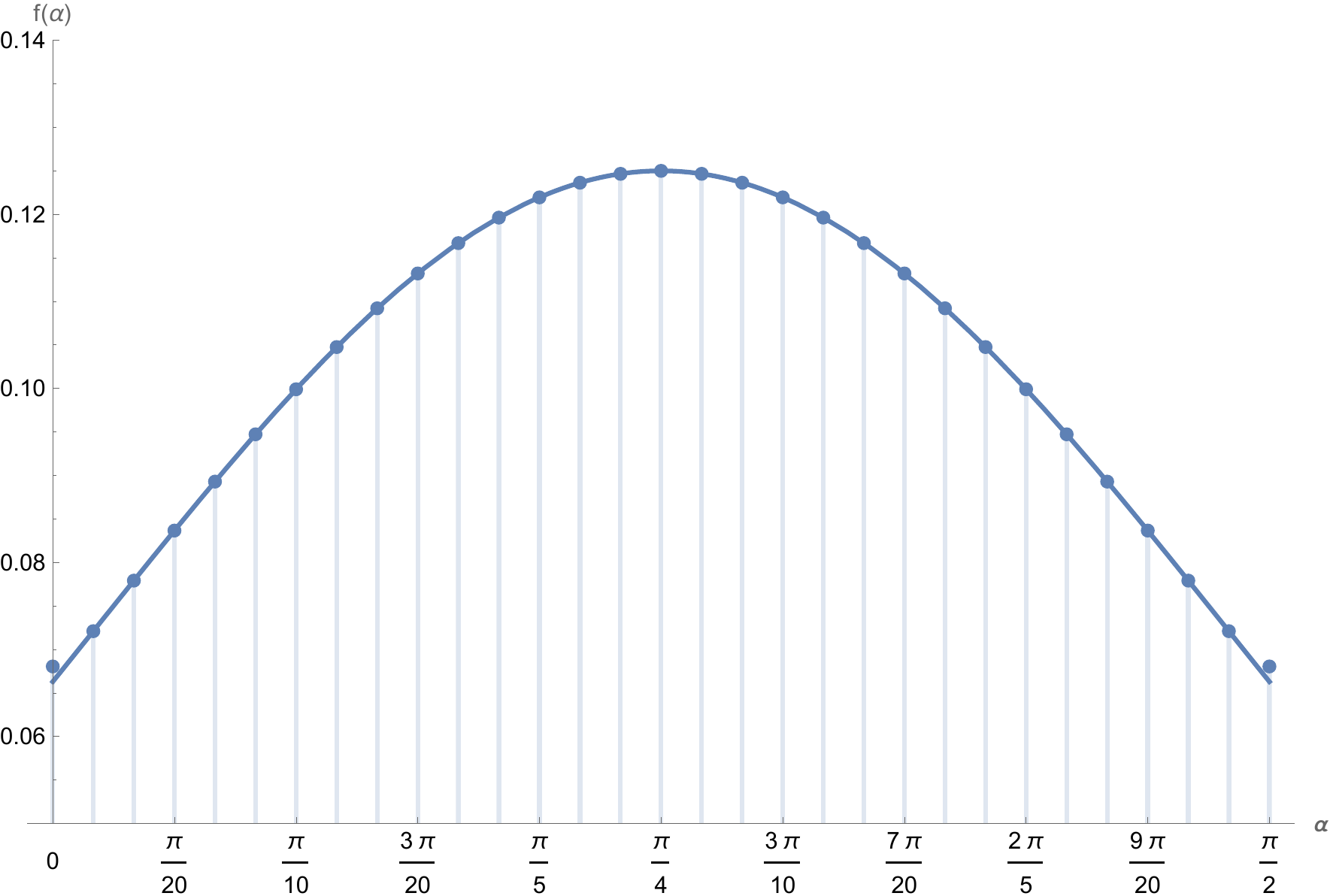}\\
			\caption{Illustration of the maximal quantum values $f(\alpha)$ in (\ref{eq:ent-6}) of uncertainty function $f$ in \eqref{eq:bellJpro}.}
			\label{fig:fa}
		\end{figure}

		\item For uncertainty function $g$, the maximal quantum value reads
		\begin{align}\label{eq:MaxViolateAdd}
			g(\alpha)=\frac{2+\sin(2\alpha)}{4},
		\end{align}
		$\alpha\in[0,\pi/2]$, which can be depicted in Fig.~\ref{fig:add}. 
		Similarly, the quantum value $g(\alpha)$ is related to the concurrence $C$ as
		\begin{align}
			g(C)=\frac{2+C}{4}.
		\end{align}

		\begin{figure}[h]
			\centering
			\includegraphics[width=8cm]{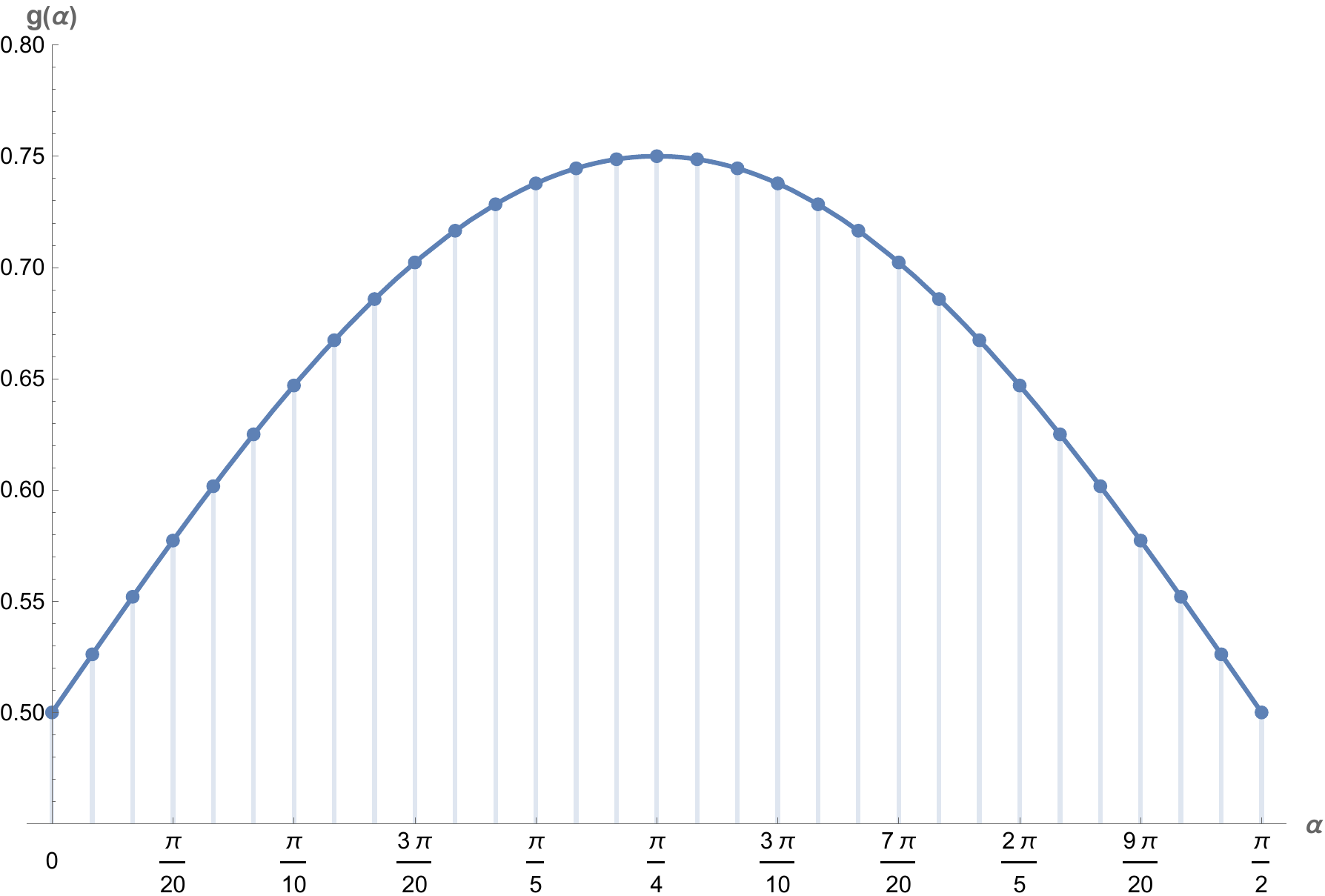}
			\caption{Depiction of the maximal quantum value $g(\alpha)$ in \eqref{eq:MaxViolateAdd} of uncertainty function $g$ in \eqref{eq:bellJadd}.}
			\label{fig:add}
		\end{figure}

		\item For uncertainty function $h$, the maximal quantum values for any 2-qubit state \eqref{eq:ent-5} can be visualized numerically in
		Fig.~\ref{fig:KMultiple}. 
		
		\begin{figure}[h]
			\centering
			\includegraphics[width=8cm]{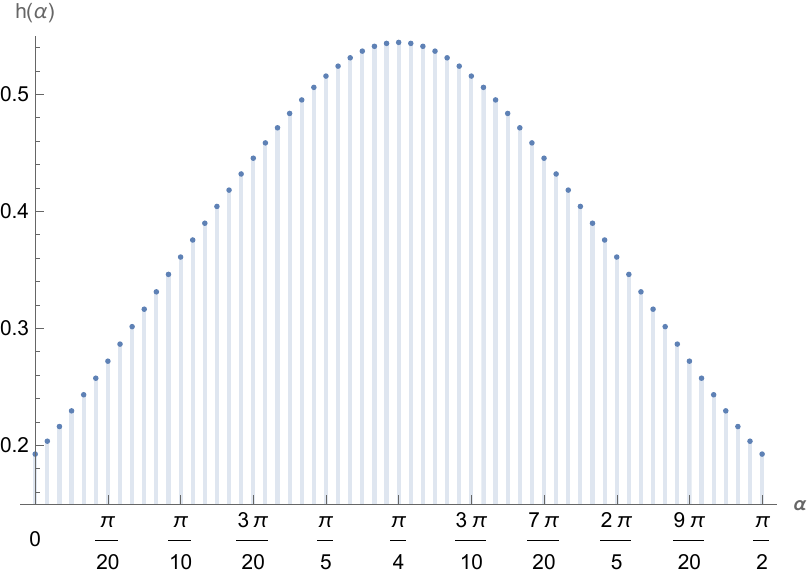}
			\caption{Depiction of the maximal quantum value $h(\alpha)$ with dotted curve of the uncertainty function $h$ in \eqref{eq:bellKpro}.}
			\label{fig:KMultiple}
		\end{figure}
		
		The maximal quantum values of uncertainty function $h$ is
		\begin{equation}
			h=-\frac{1}{2\sqrt{2} 3^{3/4}}\sqrt{h_1}+\frac{1}{128}\sqrt{h_2}*\sqrt{h_3}*\sqrt{h_4},
		\end{equation}
		with
		\begin{widetext}\textbf{}
		\begin{align*}
			&h_1=\left| \cos^3(2\alpha) [\cos(2\theta_1) + \cos(2\theta_2)] [\sin(2\theta_1) \cos\phi_1 + \sin(2\theta_2) \cos\phi_2][\sin(2\theta_1) \sin\phi_1 + \sin(2\theta_2) \sin\phi_2] \right|,\\
			&h_2=6 - \cos(4\theta_1) + 4 \cos(2\theta_1) \cos(2\theta_2) - 2 \cos(4\alpha) [\cos(2\theta_1) + \cos(2\theta_2)]^2 - \cos(4\theta_2)\\
			&\quad\quad  +8\sin(2\alpha) \sin(2\theta_1) \sin(2\theta_2) \cos(\phi_1+\phi_2),\\
			&h_3=14 + 2 \cos(4\theta_1) \cos^2\phi_1 + 16 \cos^2\theta_1 [\cos^2\theta_2 \sin(2\alpha) - \cos(4\alpha) \cos^2\phi_1 \sin^2\theta_1] \\
			&\quad\quad - \cos(2\phi_2) [1 + 16 \sin(2\alpha) \cos^2\theta_1 \sin^2\theta_2]- 16 \sin(2\alpha) \sin^2\theta_1 \sin^2\theta_2 \sin(2\phi_1) \sin(2\phi_2) \\
			&\quad\quad + 2 \cos\phi_2 \{8 \cos\phi_1 \sin^2(2\alpha) \sin(2\theta_1) \sin(2\theta_2) + \cos\phi_2 \left[\cos(4\theta_2) - 2 \cos(4\alpha) \sin^2(2\theta_2)\right]\} \\
			&\quad\quad - \cos(2\phi_1) \{1 + 16 \sin(2\alpha) \sin^2\theta_1 \left[\cos(2\theta_2) \cos^2\phi_2 + \sin^2\phi_2\right]\},\\
			&h_4=\cos(2\phi_2) [1 - 16 \sin(2\alpha) \cos^2\theta_1 \sin^2\theta_2]+ \cos(2\phi_1) \{1 - 16 \sin(2\alpha) \sin^2\theta_1 \left[\cos^2\phi_2 + \cos(2\theta_2) \sin^2\phi_2\right]\} \\
			&\quad\quad + 2 \{7 + [\cos(4\theta_1) - 2 \cos(4\alpha) \sin^2(2\theta_1)] \sin^2\phi_1 + 8 \sin^2(2\alpha) \sin(2\theta_1) \sin(2\theta_2) \sin\phi_1 \sin\phi_2 \\
			&\quad\quad+ \left[\cos(4\theta_2) - 2 \cos(4\alpha) \sin^2(2\theta_2)\right] \sin^2\phi_2 + 8 \sin(2\alpha) [-\cos^2\theta_1 \cos^2\theta_2 + \sin^2\theta_1\sin^2\theta_2 \sin(2\phi_1) \sin(2\phi_2)]. \}
		\end{align*}
		\end{widetext}
		Compared to the other three forms of the maximal values of uncertain functions ($f$, $g$ and $k$), the mathematical form of $h$ is significantly much more complex involving a root and an absolute value and the corresponding optimizing unitary operations, which leads to the absence of a clear analytical solution.  Clearly, the summation forms of the components in $g$ and $k$ are much more simpler than the product forms in $f$ and $h$. Specially, the second term in $f$  (\eqref{eq:bellJpro}) is theoretically equal to $0$, which also efficiently simplifies the expression for $f$. To sum up, we give the exact theoretical analytic dots in Fig. \ref{fig:KMultiple}.


		\item For uncertainty function $k$, the maximal quantum value reads
		\begin{align}\label{eq:MaxViolateAddS1AddS2}
			k(\alpha)=1+\sin(2\alpha),
		\end{align}
		$\alpha\in[0,\pi/2]$, which can be depicted in Fig.~\ref{fig:URofKAdd}.
		The quantum value $k(\alpha)$ is related to the concurrence $C$ as
		\begin{align}
			k(C)=1+C.
		\end{align}
		
		\begin{figure}[htbp]
			\centering
			\includegraphics[width=8cm]{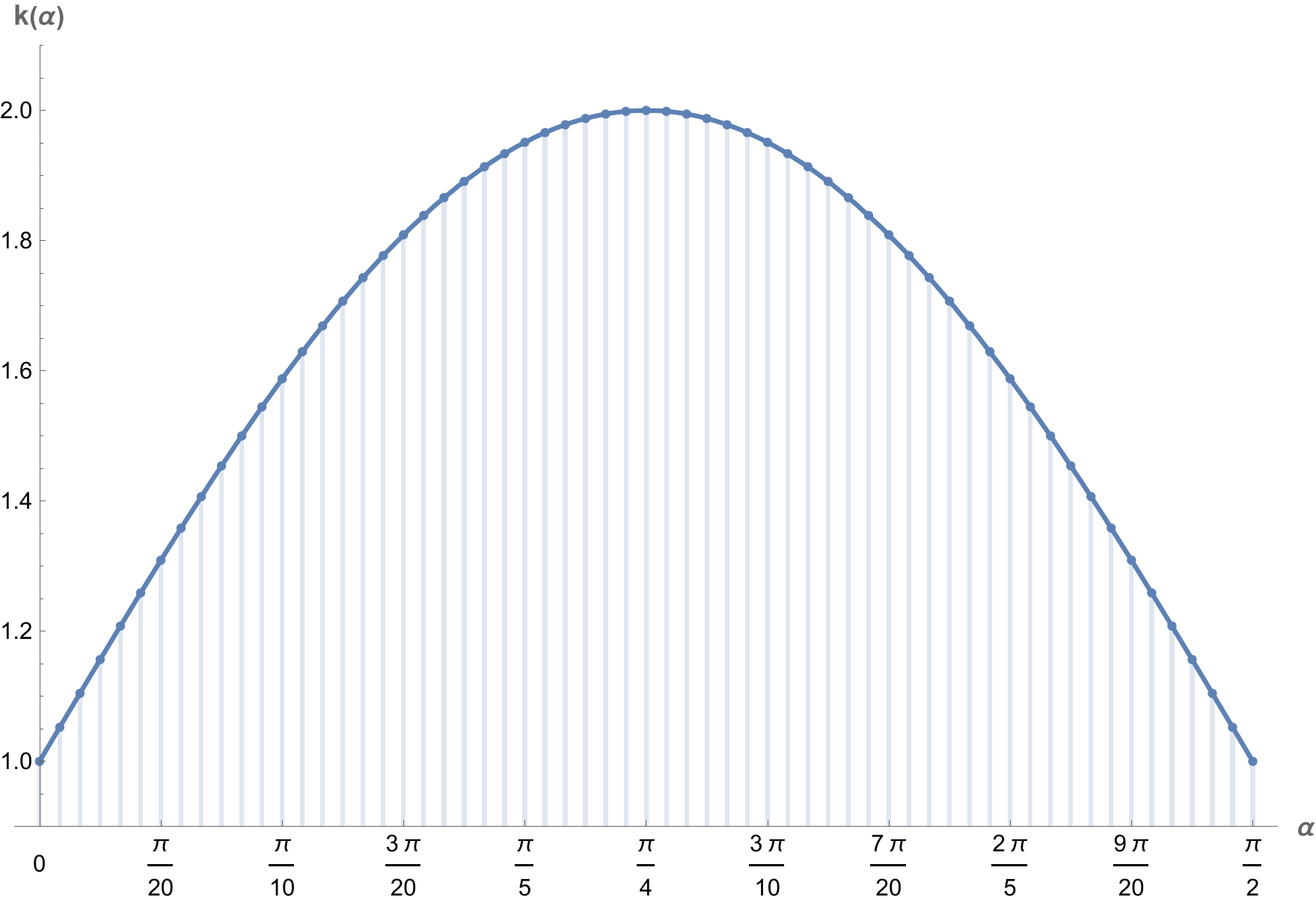}
			\caption{Depiction of the maximal quantum value $k(\alpha)$ in \eqref{eq:MaxViolateAddS1AddS2} of uncertainty function $k$ in
				\eqref{eq:bellKadd}.}\label{fig:URofKAdd}
		\end{figure}

	\end{enumerate}

	\subsection{The maximal value of four uncertainty functions with the generalized Werner state}
	
	The generalized Werner state as
	\begin{equation}\label{eq:GeneralWerner}
		\rho(\alpha,\eta)=\eta\,\ket{\Psi(\alpha)}\bra{\Psi(\alpha)}+\dfrac{1-\eta}{4}\openone\otimes\openone,
	\end{equation}
	with $\eta\in[0,1]$ the proportion of $\ket{\Psi(\alpha)}=\cos{\alpha}\ket{00}+\sin{\alpha}\ket{11}$ in the mixed state. It is not difficult to see that the generalized Werner state degenerates into 2-qubit pure state $\ket{\Psi(\alpha)}$ as \eqref{eq:ent-5} when $\eta=1$.
	
		\begin{figure}[h]
		\centering
		\includegraphics[width=9cm]{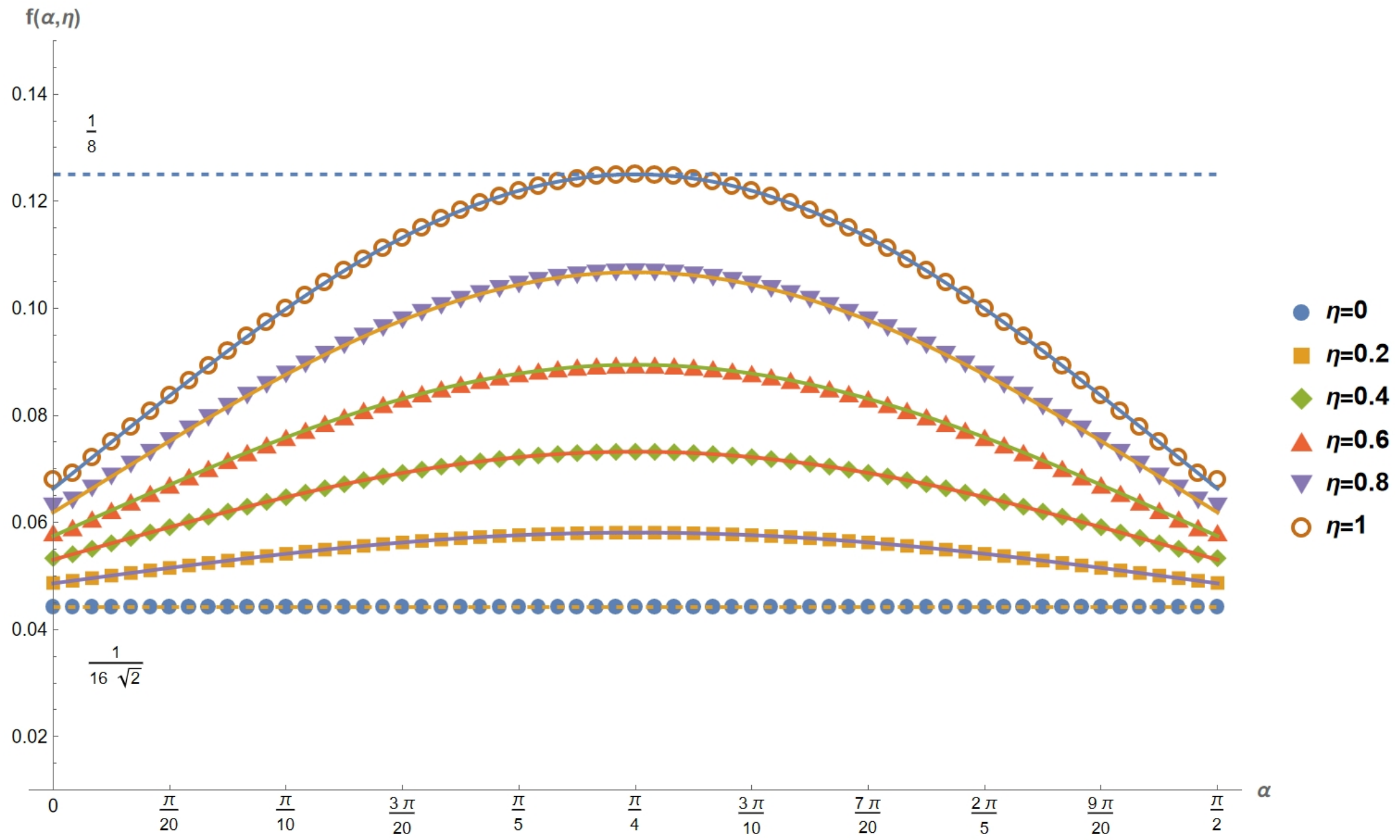}
		\caption{Depiction of the maximal quantum value $f(\alpha,\eta)$ in \eqref{eq:fQV} of uncertainty function $f$ as \eqref{eq:bellJpro} in the
			case of the generalized Werner state \eqref{eq:GeneralWerner}.}\label{fig:fQV}
	\end{figure}
	
		\begin{figure}[h]
		\centering
		\includegraphics[width=9cm]{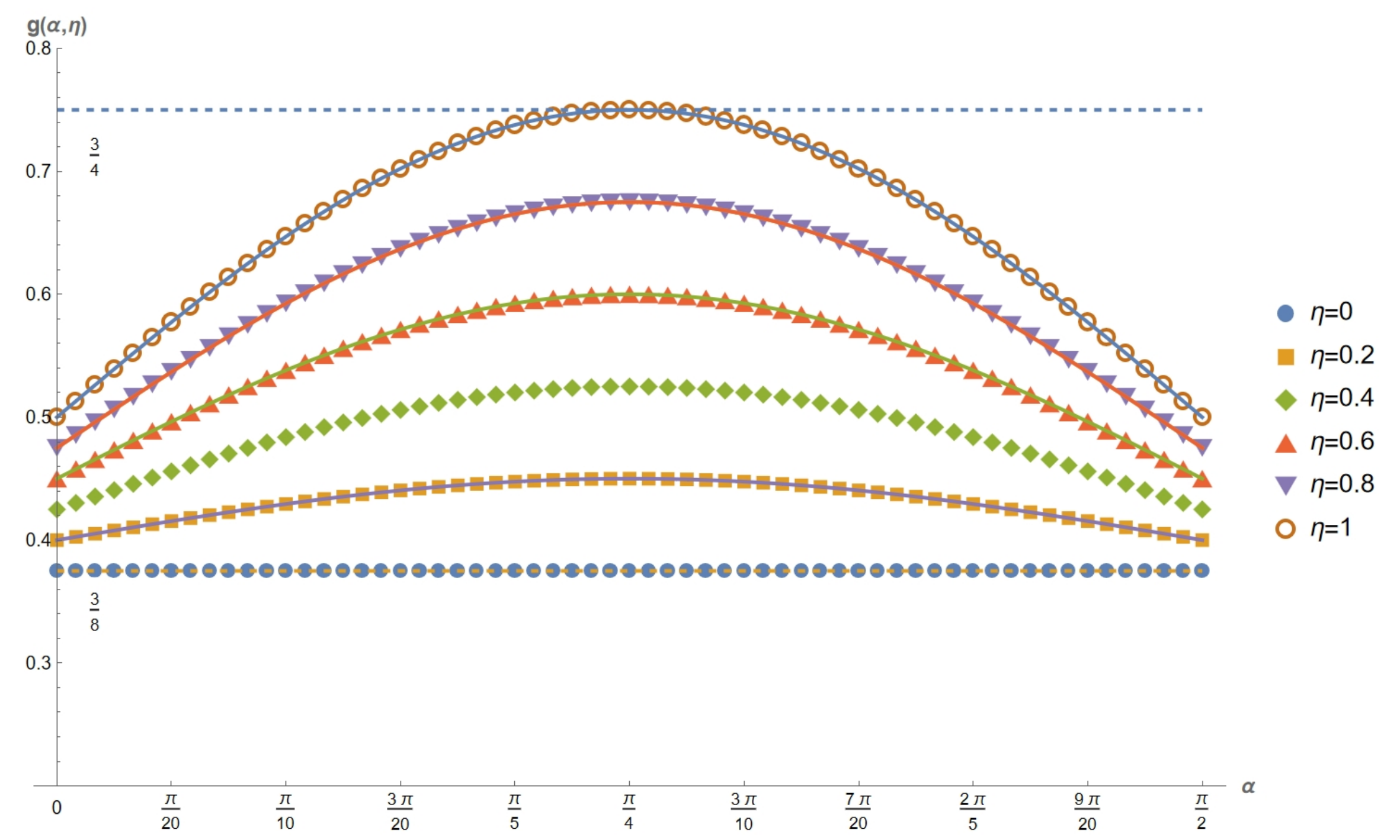}
		\caption{Depiction of the maximal quantum value $g(\alpha,\eta)$ in \eqref{eq:gQV} of uncertainty function $g$ as \eqref{eq:bellJadd} under the generalized Werner state \eqref{eq:GeneralWerner}.}\label{fig:gQV}
	\end{figure}
	
	In the following, we study the maximal quantum values of four uncertainty functions $f, h, g$, and $k$ under the generalized Werner state.
	
		\begin{figure}[htbp]
		\centering
		\includegraphics[width=8cm]{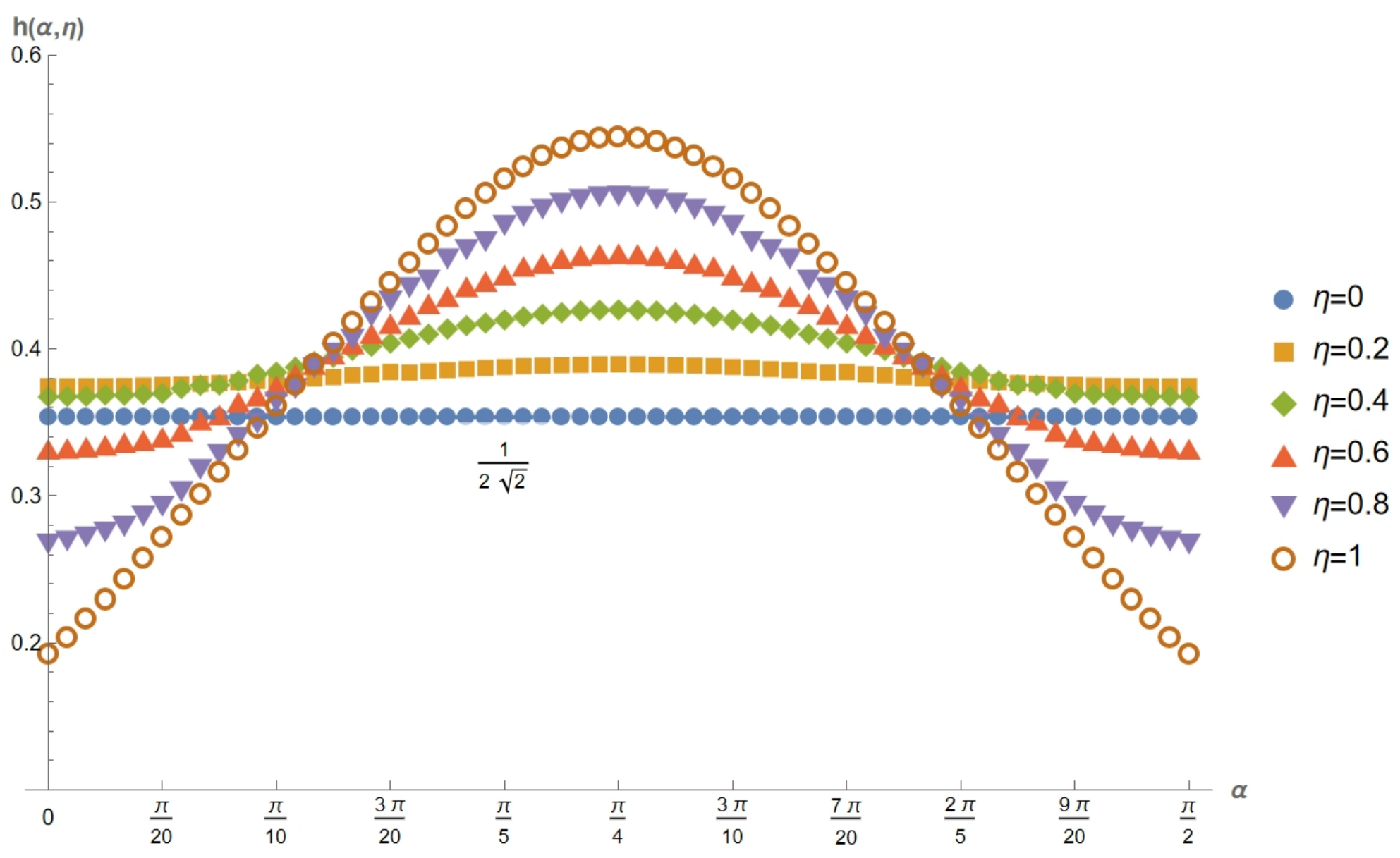}
		\caption{Illustration of the maximal quantum value $h(\alpha,\eta)$ of uncertainty function $h$ as \eqref{eq:bellKpro} for the
			generalized Werner state \eqref{eq:GeneralWerner}.}\label{fig:hQV}
	\end{figure}
	
	\begin{enumerate}
		\item For uncertainty function $f$, the maximal quantum value reads
		\begin{equation}\label{eq:fQV}
			f(\alpha,\eta)=\dfrac{\sqrt{1+\eta\sin{2\alpha}}\bigl[2+\eta(1+\sin{2\alpha})\bigr]}{32\sqrt{2}},
		\end{equation}
		see Fig.~\ref{fig:fQV}. When $\eta=1$, the maximal quantum value $f(\alpha)$ as (\ref{eq:ent-6}).

		\item For uncertainty function $g$, the maximal quantum value reads
		\begin{equation}\label{eq:gQV}
			g(\alpha,\eta)=\dfrac{3+\eta(1+2\sin{2\alpha})}{8},
		\end{equation}
		see Fig.~\ref{fig:gQV}. When $\eta=1$, the maximal quantum value $g(\alpha)$ as (\ref{eq:MaxViolateAdd}).

		\item For uncertainty function $h$, the maximal quantum value $h(\alpha,\eta)$ can be visualized numerically in Fig~\ref{fig:hQV}.

		\item For uncertainty function $k$, the maximal quantum value reads
		\begin{widetext}
			\begin{equation}\label{eq:kQV}
				k(\alpha,\eta)=\begin{cases}
					0\leq \eta\leq0.5, & \dfrac{3+\eta-\eta^2 \bigl[1+\cos(4\alpha)\bigr]}{2}, \\
					0.5<\eta\leq1, & \begin{cases}
						& \dfrac{3-\eta\bigl[1-2\sin(2\alpha)\bigr]}{2},\qquad
						\dfrac{1}{2}\arcsin\left(\dfrac{1}{\eta}-1\right)\leq\alpha\leq\dfrac{\pi}{2}-\dfrac{1}{2}\arcsin\left(\dfrac{1}{\eta}-1\right), \\
						& \dfrac{3+\eta-\eta^2 \bigl[1+\cos(4\alpha)\bigr]}{2},\quad \text{else}.
					\end{cases}
				\end{cases}
			\end{equation}
		\end{widetext}
	see Fig.~\ref{fig:kQV}. When $\eta=1$, the maximal quantum value $k(\alpha)$ as (\ref{eq:MaxViolateAddS1AddS2}).

		\begin{figure}[]
			\centering
			\includegraphics[width=8cm]{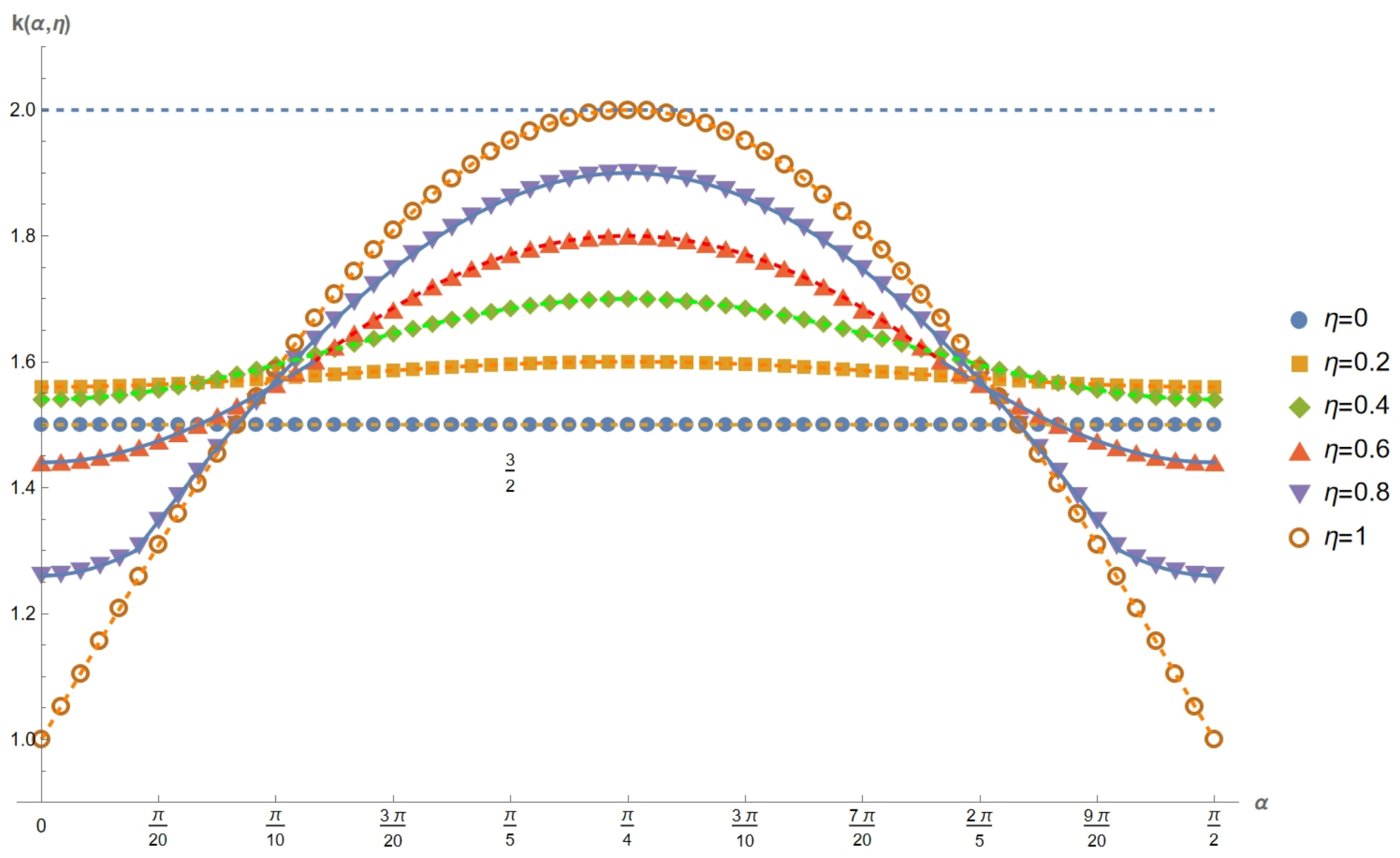}
			\caption{Depiction of the maximal quantum value $k(\alpha,\eta)$ of uncertainty function $k$ as \eqref{eq:bellKadd} in the case of the generalized Werner state \eqref{eq:GeneralWerner}.}\label{fig:kQV}
		\end{figure}

	\end{enumerate}

	%

\end{document}